\documentclass[%
 aip,
 amsmath,amssymb,
 reprint,%
]{revtex4-1}
\usepackage{amsmath,amsfonts,amssymb,amsthm,graphics,graphicx,epsfig,bbm}
\usepackage[colorlinks=true,citecolor=blue,linkcolor=blue,urlcolor=blue]{hyperref}
\usepackage[usenames]{color}
\usepackage{graphicx}
\usepackage{subfigure}
\usepackage{amsmath}
\usepackage{epsfig}
\usepackage{dcolumn}
\usepackage{bm}
\usepackage{color}
\usepackage{times}
\usepackage{epstopdf}
\usepackage{amssymb}
\usepackage{amstext}
\usepackage{latexsym}
\usepackage{hyperref}
\usepackage{amsfonts}
\usepackage{psfrag}
\usepackage{soul,xcolor}
\usepackage[normalem]{ulem}
\usepackage{dsfont}
\usepackage{txfonts}
\usepackage{physics}
\usepackage{footnote}
\usepackage{multirow}
\usepackage{appendix}
\usepackage{mathtools}
\usepackage{xspace}

\definecolor{Gray}{gray}{0.85}
\definecolor{LightCyan}{rgb}{0.88,1,1}

\newcolumntype{a}{>{\columncolor{Gray}}c}
\newcolumntype{b}{>{\columncolor{white}}c}

\newcommand{\mbf}[1]{\mathbf{#1}}
\usepackage{mathtools}

\begin{document}
\setstcolor{red}
\newtheorem{Proposition}{Proposition}[section]	
\title{Out-of-Time Ordered Correlators in Kicked Coupled Tops: Information Scrambling in Mixed Phase Space and the Role of Conserved Quantities}
\date{\today}

\author{Naga Dileep Varikuti}
\email{vndileep@physics.iitm.ac.in}
\affiliation{Department of Physics, Indian Institute of Technology Madras, Chennai, India, 600036}
	
\author{Vaibhav Madhok}
\affiliation{Department of Physics, Indian Institute of Technology Madras, Chennai, India, 600036}

\begin{abstract}
We study operator growth in a bipartite kicked coupled tops (KCT) system using out-of-time ordered correlators (OTOCs), which quantify ``information scrambling" due to chaotic dynamics and serve as a quantum analog of classical Lyapunov exponents. In the KCT system, chaos arises from the hyper-fine coupling between the spins. Due to a conservation law, the system's dynamics decompose into distinct invariant subspaces. Focusing initially on the largest subspace, we numerically verify that the OTOC growth rate aligns well with the classical Lyapunov exponent for fully chaotic dynamics. While previous studies have largely focused on scrambling in fully chaotic dynamics, works on mixed-phase space scrambling are sparse. We explore scrambling behavior in both mixed-phase space and globally chaotic dynamics. In the mixed phase space, we use Percival's conjecture to partition the eigenstates of the Floquet map into ``regular" and ``chaotic." Using these states as the initial states, we examine how their mean phase space locations affect the growth and saturation of the OTOCs. Beyond the largest subspace, we study the OTOCs across the entire system, including all other smaller subspaces. For certain initial operators, we analytically derive the OTOC saturation using random matrix theory (RMT). When the initial operators are chosen randomly from the unitarily invariant random matrix ensembles, the averaged OTOC relates to the linear entanglement entropy of the Floquet operator, as found in earlier works. For the diagonal Gaussian initial operators, we provide a simple expression for the OTOC.
\end{abstract}

\maketitle

\newtheorem{theorem}{Theorem}[section]
\newtheorem{corollary}{Corollary}[theorem]
\newtheorem{lemma}[theorem]{Lemma}
\def\endproof{\hfill$\blacksquare$}

\textbf{Quantum chaos examines how quantum systems exhibit sensitivity to initial conditions, leading to chaotic dynamical behavior. Various diagnostics, including level spacing statistics and Loschmidt echo, have been proposed to detect signatures of classical chaos in quantum systems.  The recent focus on operator growth, measured by out-of-time ordered correlators (OTOCs), stems from its connection to the maximum classical Lyapunov exponent. This study explores the OTOCs in a system of kicked coupled tops, exhibiting chaos in the classical limit and conserving total magnetization. OTOC behavior in the mixed classical phase, which remains unexplored, is studied.
We also explore the relation between system symmetries and information scrambling}

\section{Introduction}
Chaos in classical physics is closely related to non-integrability, ergodicity, complexity, entropy production, and thermalization, which are central to the study of many-body physics and classical statistical mechanics. The quantum mechanics of systems whose classical counterparts are chaotic, generally known as quantum chaos, aims to extend these ideas to the quantum domain. In classical physics, the sensitive dependence on initial conditions implies chaos. A naive generalization of classical chaos to the quantum domain fails due to the unitarity of quantum evolutions. Hence, it is necessary to find the signatures of chaos in quantum systems through various other means. Many quantities, such as level spacing statistics \cite{haake1991quantum}, entanglement entropy \cite{bandyopadhyay2002testing, bandyopadhyay2004entanglement}, out-of-time ordered correlators (OTOCs) \cite{chaos1}, and tri-partite mutual information \cite{pawan}, have emerged as powerful tools to characterize the chaos in quantum systems. OTOCs are particularly interesting due to their usefulness in characterizing various features of quantum many-body systems. Originally introduced in the theory of superconductivity \cite{larkin}, the OTOCs are being studied with renewed interest in the context of quantum many-body systems \cite{ope2, ope1, ope4, ope5, lin2018out}, quantum chaos \cite{chaos1, pawan, seshadri2018tripartite, lakshminarayan2019out, shenker2, moudgalya2019operator, manybody2, chaos2, cotler2017chaos}, many-body localization \cite{manybody3, manybody4, manybody1, huang2017out, pg2021exponential} and holographic systems\cite{shock1, shenker3}. The early time growth rate of OTOCs, in particular, is being actively studied\cite{rozenbaum2017lyapunov, prakash2020scrambling, jalabert2018semiclassical, lakshminarayan2019out, ope5, garcia2018chaos, chen2018operator, moudgalya2019operator, haehl2019classification, chen2017out, omanakuttan2019out, alonso2019out, borgonovi2019timescales, yan2020information, rozenbaum2020early, rozenbaum2019universal, lerose2020bridging}, which is a quantum counterpart of the classical Lyapunov exponent (LE).

In order to define the OTOC, consider two local Hermitian or unitary operators $A$ and $B$ acting on two disjoint local subsystems of a given system with the dimensions $d_A$ and $d_B$, respectively. Then, a function of the commutator, namely the squared commutator, is given by 
\begin{equation}\label{commutator}
C(t)=\frac{1}{2}\langle\psi| [A(t), B]^{\dagger}[A(t), B]|\psi\rangle,
\end{equation}
where $A(t)=U^{\dagger}(t)A(0)U(t)$ and $U$ represents system's time evolution operator. For the sake of simplicity and experimental feasibility, the state $|\psi\rangle$ is usually taken to be the maximally mixed state --- $\mathbb{I}/d_Ad_B$, where $d_A$ and $d_B$ denote subsystem dimensions. In this text, we consider Hermitian operators for OTOC calculations. Then, the commutator function becomes $C(t)=C_{2}(t)-C_{4}(t)$, where
\begin{eqnarray}
C_2(t)=\frac{\Tr(A^2(t)B^2)}{d_Ad_B}\quad\text{and}\quad C_4(t)=\frac{\Tr(A(t)BA(t)B)}{d_Ad_B}.
\end{eqnarray}
Here, $C_2(t)$ is a time-ordered two-point correlator, and $C_4(t)$ represents the four-point correlator function. $C_4(t)$ possesses an unusual time ordering, which is why it is referred to as OTOC (Out-of-Time-Ordered Correlator). The four-point correlator is the dominant factor driving the total commutator function $C(t)$. Hence, we use the terms OTOC and commutator function interchangeably to denote the same quantity $C(t)$.

In the semiclassical limit ($\hbar\rightarrow 0$), the Poisson bracket replaces the commutator. For $A=\hat{X}$ and $B=\hat{P}$, the semiclassical limit implies $\{ X(t), P \}^2=(\delta X(t)/\delta X(0))^2\sim e^{2\lambda t}$, where $\lambda$ denotes the maximum Lyapunov exponent (LE), which is positive for chaotic systems. The correspondence principle states that the dynamics of expectation values of observables in a quantum system follow corresponding classical equations of motion until a time known as Ehrenfest's time $(t_{\text{EF}})$  that depends on the dynamics of the system. Hence, for the quantum systems whose classical analog is chaotic, one can expect that the OTOCs grow exponentially until the Ehrenfest time $t_{\text{EF}}$ which scales with the system dimension ($\mathcal{\text{dim}}$) and the LE as $\sim \log(\mathcal{\text{dim}})/\lambda$. Though the early time growth rate of OTOC correlates well with the LE for many chaotic systems, it is worthwhile to note that the exponential growth may not always represent chaos \cite{hashimoto2020exponential, pilatowsky2020positive, pappalardi2018scrambling, hummel2019reversible, xu2020does}. Moreover, in finite temperature systems, the growth rate is bounded by $2\pi k_BT$ \cite{chaos1, murthy2019bounds}, where $k_B$ and $T$ denote the Boltzmann constant and the temperature, respectively. For $t>t_{\text{EF}}$, quantum corrections (non-zero $\hbar$) start dominating, resulting in the breakdown of the quantum-classical correspondence for the OTOC. 

Recently, OTOCs have been found to have various interesting connections to other probes of quantum chaos such as spectral form factors \cite{de2019spectral}, participation ratio \cite{borgonovi2019timescales}, universal level statistics  \cite{rozenbaum2019universal}, operator entanglement entropy \cite{styliaris2021information}, tri-partite mutual information \cite{pawan}, Loschmidt echo \cite{yan2020information}, frame potentials, and unitary designs \cite{roberts2017chaos}. OTOCs have also been investigated in systems with very few qubits and observed that the signatures of short-time exponential growth can still be found in such systems \cite{sreeram2021out}.  
Another intriguing feature of OTOC is that in multipartite systems, the growth of OTOC is related to the spreading of an initially localized operator across the system degrees of freedom. The commutator-Poisson bracket connection gives an analog of the classical separation of two trajectories with quantum mechanical operators replacing the classical phase space trajectories. Under chaotic many-body dynamics, an initially simple local operator becomes increasingly complex and will have its support grown over many system degrees of freedom, thereby making the initially localized quantum information at later times hidden from the local measurements \cite{pawan, hayden2007black, moudgalya2019operator, ope1}. As the operator evolves in the Heisenberg picture under chaotic dynamics, it will start to become incompatible with any other operator with which it commutes initially. Bi-partite systems are the simplest multipartite settings to study such a scenario. In this paper, we study the dynamics of OTOCs in the kicked coupled tops (KCT), a bipartite system with a well-defined classical limit. In prior studies, many authors have considered systems of two (or more) degrees of freedom with time-dependent Hamiltonians of the form $H(t) = H_1(t) + H_2(t) + H_{12}(t)$, where the classical dynamics generated by $H_1$ and $H_2$ can exhibit chaos for each degree of freedom separately, and the coupling interaction $H_{12}$ between them can be independently varied. For such systems, the chaoticity parameter and coupling constant play different roles. To this effect, a system of coupled kicked rotors has been previously studied \cite{prakash2020scrambling, prakash2019out} where two kicked rotors that independently exhibit chaos were weakly coupled. In this work, we study a system where the coupling strength is the chaoticity parameter, and chaos occurs due to this mechanism rather than separately in the two systems.

In this work, our primary goal is to examine operator growth as quantified by the OTOCs in the system of kicked coupled tops. In the context of classical-quantum correspondence, works thus far have largely focussed on the OTOCs in quantum systems with chaotic classical limits. However, the behavior of operator growth in the mixed-phase space remains poorly understood with only a few studies \cite{mondal2021dynamical, notenson2023classical, bergamasco2019out, roy2021entanglement, kidd2021saddle, kidd2020thermalization}. This prompts us to study the role of the mixed-phase space dynamics on operator growth along with the fully chaotic dynamics. Studies in the context of mixed-phase space are interesting for the following reasons: Firstly, the random matrix theory (RMT) has been successful in characterizing the quantum systems that are chaotic in the classical limit. RMT can accurately explain level spacing distributions as well as other statistical properties like the saturation value of entanglement for a globally chaotic system. Furthermore, the regular systems (with Poisson statistics) can be well described using the diagonal unitaries with the diagonal elements chosen uniformly at random from the unit complex circle. However, in the case of mixed systems, the RMT can only capture the universal properties associated with the ratio of the phase space volumes occupied by the regular and chaotic regions \cite{rosenzweig1960repulsion, berry1984semiclassical, prosen1994semiclassical, robnik2000topics, kravtsov2015random}. Secondly, classical systems in mixed-phase space enforce a perfect separation between chaotic and regular trajectories. While the former live in the chaotic sea and have a positive Lyapunov exponent, the latter reside in the regular islands and show no chaos. However, quantum mechanically, the time-evolved operators corresponding to the mixed phase space will have support over both regular and chaotic regions of the phase space, making direct quantum-classical correspondence extremely challenging.

The works thus far have studied the mixed phase space scrambling in a few different models such as standard map \cite{notenson2023classical}, coupled cat maps \cite{bergamasco2019out}, coupled kicked rotors \cite{prakash2020scrambling}, and Bose-Hubbard models \cite{kidd2020thermalization, kidd2021saddle}. 
It has been argued that in these systems, the short-time growth provides an ambiguous indicator of transition to chaos \cite{kidd2021saddle}. This is partly because of the ambiguity associated with the states lie at the boundaries between the regular and chaotic regions. These states despite being regular, can show characteristics (such as entanglement entropy) similar to the chaotic states \cite{lombardi2011entanglement, madhok2015comment}. Nevertheless, the long-time dynamics of the OTOCs provide a much clearer indicator of chaos in these cases \cite{garcia2018chaos}. Moreover, in fully chaotic systems, the OTOCs approach saturation exponentially, while mixed phase space leads to a multi-step relaxation of OTOCs towards saturation \cite{notenson2023classical}. In this work, to study the scrambling in mixed phase space, we approach a different route. In particular, we invoke Percival's conjecture \cite{percival1973regular} and partition the eigenstates of the Floquet map into ``regular" and ``chaotic" and examine the behavior of OTOCs in those states. 

Apart from being chaotic, the system we consider also displays a conservation law. The conserved quantities have been shown to slow the information scrambling in various systems \cite{chen2020quantum, ope4, ope5, kudler2021information, cheng2021scrambling, balachandran2021eth}. These works typically explore random unitary circuits or spin chains as physical models. These models, however, lack a smooth classical limit. In the latter part of this work, we explore how the conserved quantity constrains the growth of OTOCs as the system approaches a classical limit, highlighting the interplay between symmetries and chaos.

This paper is structured as follows. In section \ref{Model}, we review some of the details of the KCT model. Section \ref{Information scrambling in the KCT model} is devoted to analyzing information scrambling in the KCT model. In section \ref{OTOC in the largest invariant subspace} and \ref{Scrambling in the mixed phase space}, we examine scrambling in the largest subspace for the fully chaotic and mixed phase space regimes, respectively. In section \ref{the IzJzOTOC} and \ref{the IxJxOTOC}, we consider the entire system KCT system and study the scrambling for two different initial operator choices. Section \ref{Scrambling, operator entanglement and coherence} is devoted to the cases of random operators, wherein we show that the OTOC for the random operators is connected to the operator entanglement and coherent generating power of the time evolution operator. Finally, we conclude the text in section \ref{SUMMARY AND DISCUSSIONS}. 

\section{Model: Kicked coupled top}\label{Model}
The kicked coupled top (KCT) model, originally introduced in Ref. \cite{trail2008entanglement}, is governed by the following Hamiltonian:
\begin{equation}
H=\dfrac{\alpha}{\sqrt{|\mathbf{I}||\mathbf{J}|}} \mathbf{I.J} +\beta\sum_{n=-\infty}^{\infty}\delta(t-n\tau) J_{z},
\end{equation}
where $\alpha$ and $\beta$ denote coupling strength and kicking strength, respectively. For simplicity, here and throughout, we take $|\mathbf{I}|=|\mathbf{J}|=J$. The system evolution can be understood as alternating rotations of $\mathbf{J}$ around $z$-axis, followed by a precession of $\mathbf{I}$ and $\mathbf{J}$ about $\mathbf{F=I+J}$ by an angle proportional to $\alpha |\mathbf{F}|$. Total magnetization along $z$-axis ($F_z=\hat{I}_{z}+\hat{J}_{z}$) is the only known constant motion in this system. The absence of enough conserved quantities makes the system non-integrable and also chaotic when $\alpha$ is sufficiently large. 
We note that a coupled top system has been studied earlier \cite{feingold1983regular}, and a time-independent variant of this model has also been studied \cite{fan2017quantum}. 
In our work, the motivation for choosing the particular system is that this Hamiltonian describes the hyperfine interaction between nuclear spin and total electron angular momentum with a magnetic field that has a negligible effect on the nucleus. While not extending deeply into the semiclassical regime, this realization remains suitable for investigating non-trivial mesoscopic regimes in large atoms with heavy nuclei and many valence shell electrons \cite{trail2008entanglement}. In contrast, a recent study \cite{mondal2021dynamical} on coupled tops had a Hamiltonian that does not have an isotropic interaction. Moreover, the external field in their model acts on both spins, and their main objective is to use ``Fidelity OTOCs" (FOTOCS) to detect scars.

\begin{figure}
\includegraphics[scale=0.43]{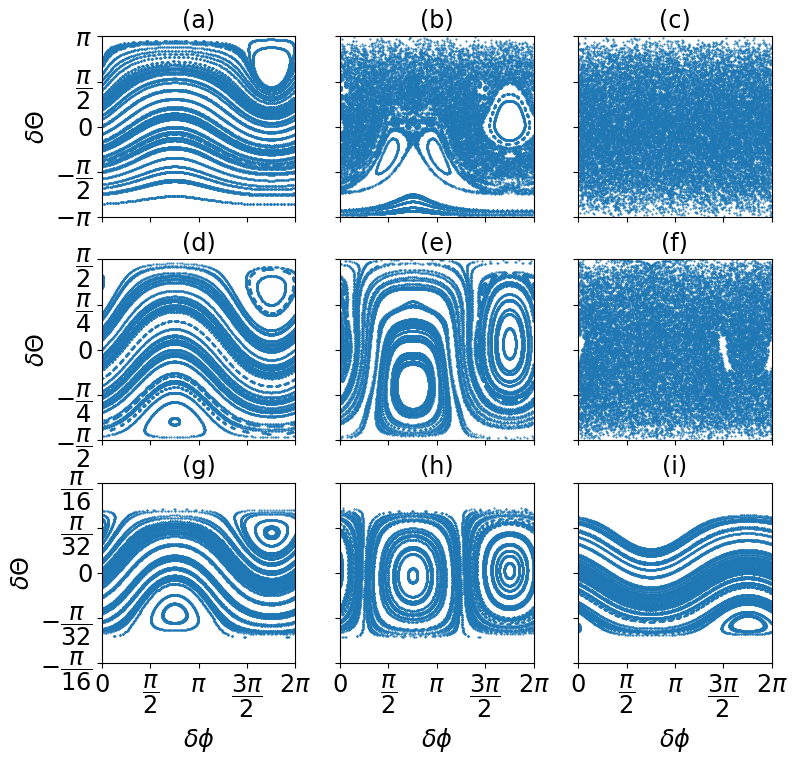}
\caption{\label{fig1} Poincar\'e\xspace surface of sections corresponding to different coupling strengths (along the rows) and different sectors of $F_z$ (along the columns). The kicking strength $\beta$ is kept constant at $\pi/2$. Panels (a)-(c) illustrate the case where $F_z=0$. Along the row, the coupling strengths are $\alpha=1/2$, $3/2$, and $6$ from left to right, respectively. In panels (d)-(f), $F_z=1$, while panels (g)-(i) illustrate $F_z=1.99$. For $\alpha=1/2$, the phase space in the sectors close to $F_z=0$ remains mostly regular across all $F_z$ sectors. However, $\alpha=3/2$ and $\alpha=6$ correspond to mixed and fully chaotic phase spaces. The phase space maintains regularity for all the couplings in the smaller sectors that are close to $|F_z|=2$.}
\end{figure}

Initially, the classical KCT model exhibits six degrees of freedom, constrained by $|\mathbf{I}|=|\mathbf{J}|=J$. This constraint results in four effective degrees of freedom. Conservation of $F_z$ further reduces the phase space from four to three dimensions. Additionally, fixing $F_z$ can effectively reduce the phase space to two dimensions. To actually visualize this, we write the Hamiltonian in terms of canonical conjugate pairs of sum and difference coordinates given by $(F_{z}=I_{z}+J_{z}, \bar{\phi}=\phi_{I}+\phi_{J})$ and $(\delta F_{z}=I_{z}-J_{z}, \delta \phi = \phi_{I}-\phi_{J})$. Then, the Hamiltonian becomes
\begin{eqnarray}\label{classical_eq}
H&=&\alpha\left[I_{z}J_{z}+|\mathbf{I}||\mathbf{J}|(\sin(\phi_{I})\sin(\phi_{J})+\cos(\phi_{I})\cos(\phi_{J}))\right]\nonumber\\
&&+\beta \sum_{n=-\infty}^{\infty}\delta(t-n\tau) J_{z}\nonumber\\
&=&\alpha\left(\frac{F_z^2-\delta F_z^2}{4}+ |\mathbf{I}||\mathbf{J}|\cos(\delta\phi)\right)\nonumber\\
&&\hspace{2cm}+\beta\sum_{n=-\infty}^{\infty}\delta(t-n\tau)\left( \frac{F_{z}-\delta F_{z}}{2} \right), 
\end{eqnarray}
where we have defined $I_x=|\mathbf{I}|\cos(\phi_{I})$ and $I_y=|\mathbf{I}|\sin(\phi_{I})$ and similarly, $J_x=|\mathbf{J}|\cos(\phi_{J})$ and $J_y=|\mathbf{J}|\sin(\phi_{J})$.
Since $F_z$ is conserved, $\overline{\phi}$ remains a cyclic coordinate and does not appear in the Hamiltonian. Therefore, to examine an invariant Poincar\'e\xspace section corresponding to a constant $F_z$, we only need the variables $(\delta F_z, \delta\phi)$. For $F_z=0$, the corresponding Hamiltonian is
\begin{equation}\label{fzzeoHam}
H={\alpha}\left[\frac{-\delta F_{z}^2}{4}+|\mathbf{I}||\mathbf{J}| \cos(\delta \phi)\right]-\beta\sum_{n=-\infty}^{\infty}\delta(t-n\tau) \frac{\delta F_{z}}{2} .
\end{equation}
By writing $\delta F_z= \cos\theta_I -\cos\theta_J =\sin(\delta\theta/2)\sin(\overline{\theta}/2)$, where $\overline{\theta}=\theta_I+\theta_J=\pi$ and $\delta\theta=\theta_I-\theta_J$, the phase space can be visualized in $(\delta\theta, \delta\phi)$ variables for various initial conditions.

\begin{figure}
\includegraphics[scale=0.375]{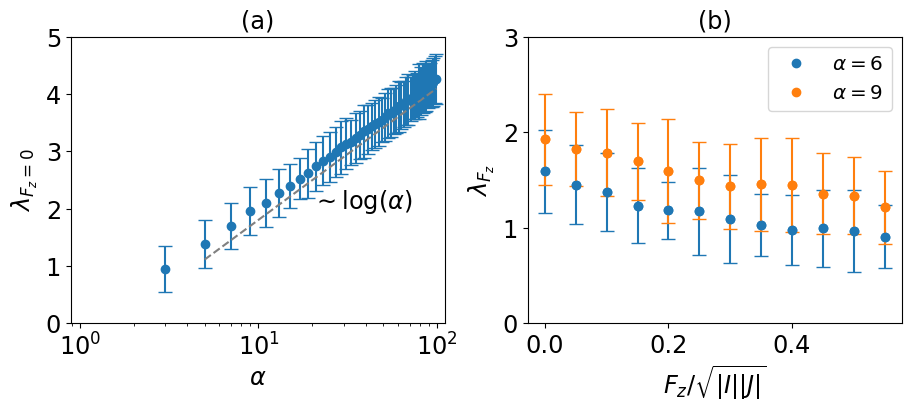}
\caption{\label{lyap} (a) Illustration of the maximum Lyapunov exponent vs. coupling strength $\alpha$ in the $F_z=0$ sector averaged over $10^3$ initial conditions. The plot suggests that the average Lyapunov exponent in the largest invariant subspace ($\lambda_{F_z=0}$) follows an approximately logarithmic law. (b) The Lyapunov exponent is calculated for different values of $F_z$. The interaction strength and the kicking parameter are kept fixed at $6$ and $\pi/2$, respectively. The error bars represent the standard deviation of the maximum Lyapunov exponent.}
\end{figure}

By noting that the rotations of the classical vectors can be implemented by the SO($3$) operators, the phase space of the KCT model can be easily visualized. To be specific, the evolution of the classical angular momentum vectors $\mathbf{I}$ and $\mathbf{J}$ can be written as follows:    
\begin{eqnarray}\label{In}
\mathbf{I}(n+1) = \exp\left\{ \alpha\left( F_x\hat{L}_x+F_y\hat{L}_{y}+F_zL_z \right) \right\}\mathbf{I}(n)  
\end{eqnarray}
and 
\begin{eqnarray}\label{Jn}
\mathbf{J}(n+1)=\exp\left\{ \beta L_z \right\} \exp\left\{ \alpha\left( F_x\hat{L}_x+F_y\hat{L}_{y}+F_zL_z \right) \right\}\mathbf{J}(n), 
\end{eqnarray}
where $L_{x, y, z}$ are the generators of the SO$(3)$ group, and the sum of the two vectors is given by $\mathbf{F}=\mathbf{I}+\mathbf{J}=[F_x, F_y, F_z]$. For convenience, in the above equations, we have normalized the classical vectors to $1$. The Poincar\'e\xspace sections corresponding to three different $F_z$ sectors, namely, $F_z=0$, $1$ and $1.99$ (along the columns) are shown in Fig. (\ref{fig1}) for a few different $\alpha$ values along the rows. In the absence of coupling, the system remains integrable. In the largest sector ($F_z=0$), we observe that the chaos slowly builds up with the increase in $\alpha$. However, the sectors far from $F_z=0$ show regular dynamics even when the $\alpha$ is large as evident from Fig. \ref{fig1}g-\ref{fig1}i. Figure (\ref{lyap}a) shows the illustration of maximum LE (averaged over many initial conditions) vs. $\alpha$ calculated using Benettin's algorithm \cite{benettin1980lyapunov, kuznetsov2012hyperbolic} in $F_z=0$ section. The plot suggests that the growth of the average LE versus the coupling strength is logarithmic --- $\lambda\sim\log(\alpha)$. Interestingly, this is very similar to the analytical calculation of LE done for the kicked top in Ref. \cite{PhysRevE.56.5189}. Moreover, as $F_z$ increases, the area of the corresponding phase space sector decreases. As a result, the largest Lyapunov exponent $\lambda_{F_z}$, which roughly scales like $\sim\log(\text{Area})$, decreases. Fig. (\ref{lyap}b) illustrates the same for two different $\alpha$ values. We observe that $\lambda_{F_z}$ decreases linearly with $F_z$. Near $|F_z|=2J$, the area tends to vanish, leading to highly localized dynamics. 

In the quantum regime, the evolution of the system is given by the following Floquet operator:
\begin{eqnarray}\label{KCT}
U=\exp\left\{\dfrac{-i\alpha}{\sqrt{IJ}} \mathbf{I.J}\right\} \exp\left\{-i\beta \hat{J}_{z}\right\},
\end{eqnarray}
where $\mathbf{I.J}=\hat{I}_x\hat{J}_x+\hat{I}_y\hat{J}_y+\hat{I}_z\hat{J}_z$
The quantum version of the KCT is also non-integrable due to insufficient conservation laws. In the uncoupled representation, the elements of $U$ can be written as follows:
\begin{eqnarray}\label{floq}
U_{m_1m_2, m_3m_4}=\sum_{F=|M|}^{2J}e^{-i\frac{\alpha}{2\sqrt{|I||J|}}F(F+1)}e^{-i\beta m_4}C_{I,m_1;J, m_2}^{F, M}C_{I,m_3;J, m_4}^{F, M},
\end{eqnarray}
where $C_{I, m_i;J, m_j}^{F, M}=\langle F, M|I, m_i;J, m_j\rangle$ denote Clebsch-Gordan coefficients. Similar to its classical counterpart, $U$ also admits a decomposition into various invariant subspaces characterized by the quantum number $F_z$ --- $U=\oplus_{F_z}U_{F_z}$, where $F_z$ runs from $-2J$ to $2J$. As a result, $U_{m_1m_2, m_3m_4}=0$ whenever $|m_1+m_2|\neq|m_3+m_4|$. Moreover, the dimension of each subspace is related to $F_z$ as $d=2J+1-|F_z|$.

An interesting feature of quantum chaos is its connections with random matrix theory (RMT). Dyson introduced three random unitary ensembles \cite{dyson1962threefold}, namely circular unitary ensemble (CUE), circular orthogonal ensemble (COE), and circular symplectic ensemble (CSE) based on their properties under time-reversal operation. If a system exhibits chaos in the classical limit ($\hbar \rightarrow 0$), the spectral statistics of its evolution align with one of the three random unitary ensembles. In the present case, we consider the generalized time-reversal operation and see that the system under study is invariant under the time-reversal operation.
\begin{equation}
\label{Eq:Reversal}
T=e^{i \beta \hat{J}_{z}} K,
\end{equation}
where $K$ is complex conjugation in the uncoupled product basis.  Since both $\hat{I}_{y}$ and $\hat{J}_{y}$ change sign under conjugation, while the $x$ and $z$ components of the angular momentum remain unaffected,
\begin{equation}
K \hat{J}_{z} K = \hat{J}_{z}; \hspace{1 pc} K(\mbf{I} \cdot \mbf{J})K = \mbf{I} \cdot \mbf{J}.
\end{equation}

Hence,
\begin{eqnarray}
T U_{\tau} T^{-1} &=& \left( e^{i \beta \hat{J}_{z}} K \right) \left( e^{-i\tilde{\alpha} \mbf{I} \cdot \mbf{J}}  e^{-i \beta \hat{J}_{z}}  \right) \left( K e^{-i \beta \hat{J}_{z}} \right)  \\
&=&  e^{i \beta \hat{J}_{z}} \left( e^{i \tilde{\alpha} \mbf{I}\cdot \mbf{J}} e^{i \beta \hat{J}_{z}}\right) e^{-i \beta \hat{J}_{z}} \nonumber\\
&=& e^{i \beta \hat{J}_{z}}e^{i \tilde{\alpha}\mbf{I}\cdot \mbf{J}}= U_{\tau}^{\dagger}. \nonumber
\end{eqnarray}
As a result, the dynamics are invariant under time reversal operation. Moreover, $T^2=1$. Since no additional discrete symmetries are present, COE is the appropriate unitary ensemble for the KCT model.

\section{Information scrambling in the KCT model}\label{Information scrambling in the KCT model}
Recall from the previous section that the KCT model conserves $\hat{F}_z = \hat{I}_z + \hat{J}_z$, i.e., $[U, \hat{F_z}]=0$. The conservation law causes the decomposition of the dynamics into invariant subspaces. Among these subspaces, the largest one typically remains dominant. Hence, scrambling in this system is mainly driven by the largest subspace dynamics, but the regular dynamics in smaller subspaces somewhat offset it. In the following, we will examine scrambling in the largest subspace, emphasizing OTOC dynamics in the chaotic and mixed phase space regimes. Afterward, we will consider the entire system and analyze the OTOC dynamics for two pairs of initial operators, namely, $(\hat{I}_{z}, \hat{J}_{z})$ --- where both operators commute with the total magnetization operator $\hat{F}_z$, and $(\hat{I}_{x}, \hat{J}_{x})$ --- where neither operator commutes with $\hat{F}_z$.

\subsection{OTOC in the largest invariant subspace}\label{OTOC in the largest invariant subspace}
\subsubsection{Short-time growth}
Unlike smaller subspaces ($|F_z|\gg 0$), at sufficiently strong coupling $(\alpha\gtrsim 6)$, the subspaces close to $F_z=0$ are predominantly chaotic. Here, we numerically study the OTOC for a pair of operators acting exclusively on the largest invariant subspace ($F_z=0$) of the KCT model. 
The corresponding Floquet evolution in this subspace can be written as follows:
\begin{equation}\label{floqsubspace}
U_{F_z=0}=\sum_{F=0}^{2J}\sum_{m_1, m_2}e^{-i(\frac{\alpha}{2J}F(F+1)+\beta m_2)}C_{I,m_1;J, -m_1}^{F, 0}C_{I,m_2;J, -m_2}^{F, 0}|m_1\rangle\langle m_2|.
\end{equation}
Here, the indices $m_1$ and $m_2$ run from $-J$ to $J$, implying that the unitary acts on $(2J+1)$-dimensional Hilbert space.

\begin{figure}
\includegraphics[scale=0.375]{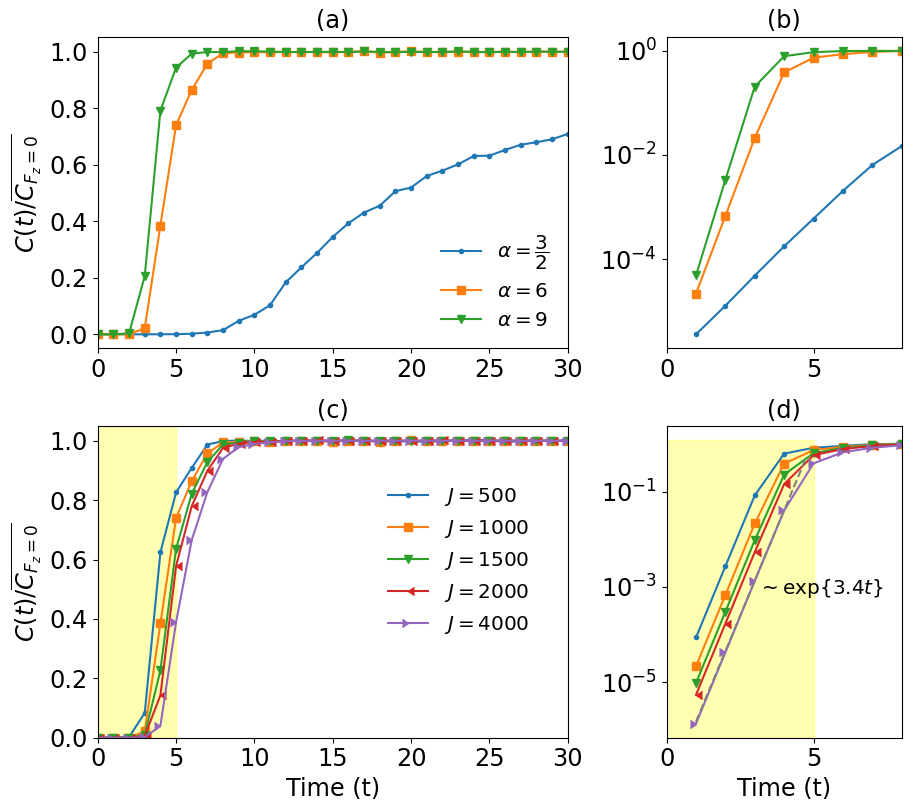}
\caption{\label{largestotoc} Illustration of the OTOC for different $\alpha$ and $J$ values in the largest invariant subspace. The initial operators are $A=B=\hat{S}_z$, the generator of rotations around $z$-axis. These operators act exclusively on the largest subspace $F_z=0$. The initial state is $\rho=\mathbb{I}/(2J+1)$, the maximally mixed state. In panel (a), the OTOC is shown for three different couplings, namely, $\alpha=3/2$, $6$, and $9$. The magnitude of $J$ is fixed at $1000$. Panel (b) shows corresponding early-time exponential growth on a semi-log plot. The plot indicates that the OTOC growth rate increases with an increase in $\alpha$. Panels (c) and (d) demonstrate OTOC growth for varying $J$ values while keeping $\alpha=6$. The plots are normalized by dividing the OTOC $C_{F_z=0}(t)$ with $\overline{C_{F_z=0}}$, the infinite time average of the OTOC.} 
\end{figure}

For the OTOC calculations, we take the initial operators $A=B=\hat{S}_z$, where $S_z$, the generator of rotation along $z$-axis --- $\hat{S}_{z}=\sum_{m=-J}^{J}m|m\rangle\langle m|$, acts non-trivially on the $F_z=0$ subspace. Then, the OTOC is given by 
\begin{eqnarray}
 C_{F_z=0}(t)=\dfrac{1}{2J+1}\left[ \Tr\left( \hat{S}^2_z(t)\hat{S}^2_{z} \right)-\Tr\left( \hat{S}_{z}(t)\hat{S}_z\hat{S}_z(t)\hat{S}_z \right) \right],  
\end{eqnarray}
where $S_z(t)$ denotes the Heisenberg evolution of $S_z$ under the dynamics given in Eq. (\ref{floqsubspace}). Here, we are fixing the maximally mixed state $\rho=\mathbb{I}/(2J+1)$ as the initial state for the OTOC calculations. In the classical limit, for chaotic systems, the Ehrenfest time is expected to scale as $t_{\text{EF}}\sim\log(\text{dim})/\lambda_{\text{cl}}$, where $\lambda_{\text{cl}}$ is the corresponding classical Lyapunov exponent. Within this timescale, $C_{F_z=0}(t)$ shows an exponential growth if $\alpha$ is sufficiently large. The corresponding numerical results are shown in Fig. \ref{largestotoc}. Figure \ref{largestotoc}a illustrates the operator growth for different values of $\alpha$ by plotting the normalized OTOC --- $C_{F_z=0}(t)/\overline{C_{F_z=0}}$ versus time. Here, $\overline{C_{F_z=0}}$ denotes infinite time average of $C_{F_{z}=0}(t)$ and is given by
\begin{eqnarray}
\overline{C_{F_z=0}}&=&\lim_{t\rightarrow\infty}\dfrac{1}{t}\int_{0}^{t}C_{F_z=0}(s)ds\nonumber\\
&=&\dfrac{1}{2J+1}\left[\sum_{m}[A^2]_{m m}[B^2]_{mm}-[A_{mm}]^2[B_{mm}]^2\right.\nonumber\\
&&\left.-\sum_{m\neq n}\left(A_{m m}A_{n n}B_{m n}B_{nm}+A_{mn}A_{n m}B_{m m}B_{nn}\right)\right],
\end{eqnarray}
where $A_{mn}=\langle E_m|A|E_n\rangle$ etc. and $U_{F_z=0}=\sum_{n}e^{-i\phi_{n}}|E_n\rangle\langle E_n|$ denotes the eigen-decomposition of the subspace Floquet operator. In the figure, the angular momentum is kept fixed at $J=1000$. The figure shows that the growth rate increases with $\alpha$. Moreover, for $\alpha=3/2$, the classical phase space has a mix of regular islands and the chaotic sea [see Fig. \ref{fig1}b]. This results in a slower OTOC growth than the fully chaotic cases (corresponding to $\alpha=6$ and $9$), as illustrated in figure \ref{largestotoc}a. We elaborate more on the mixed-phase space OTOC dynamics in the next subsection. Figure \ref{largestotoc}b demonstrates the corresponding early-time exponential growths of the OTOCs plotted in \ref{largestotoc}a. In Fig. \ref{largestotoc}c and \ref{largestotoc}d, we take the fully chaotic case by fixing $\alpha=6$ and contrast the OTOC growth for different dimensions. The corresponding numerical results yield a quantum exponent of $\lambda_{\text{quant}}\approx 1.67$ for $J=4000$, closely matching the classical LE, $\lambda_{\text{cl}}\approx 1.55$, indicating a good classical-quantum correspondence. The quantum exponent is extracted by fitting the first five data points ($0\leq t<4$) in the curve to an exponential scale. A detailed presentation of the quantum exponents for different $J$-values is given in the table. \ref{table1}.
\begin{figure}
\includegraphics[scale=0.375]{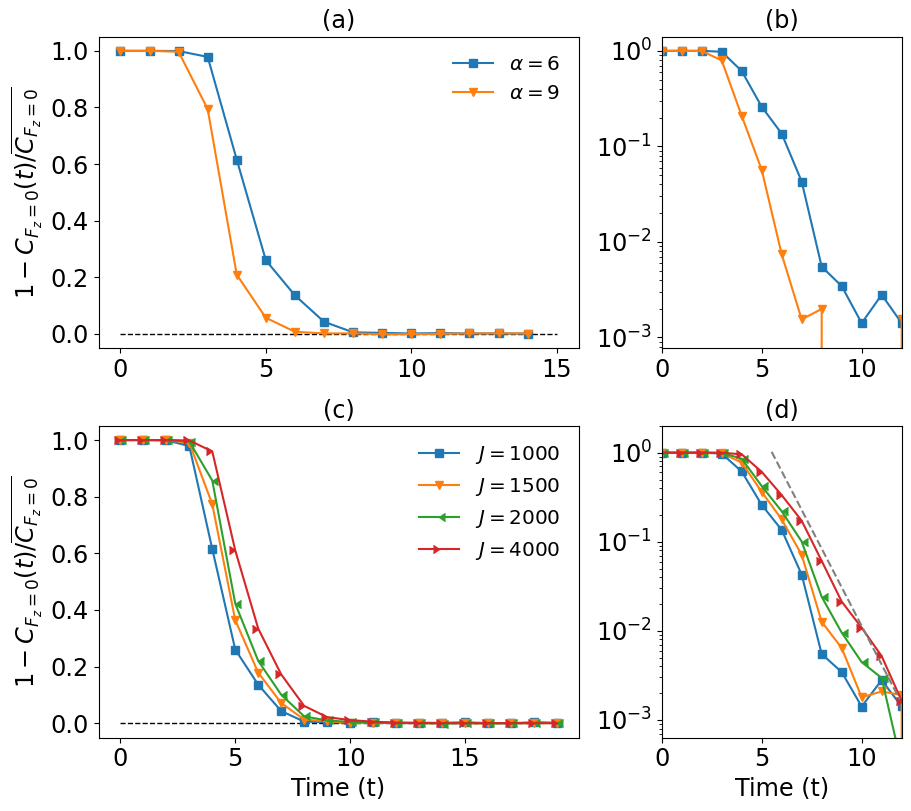}
\caption{\label{largest_relax} Relaxation dynamics of the OTOC as probed by the quantity $1-C_{F_z=0}(t)/\overline{C_{F_z=0}}$. The initial operators are the same as before: $A=B=\hat{S}_z$. Panel (a) and (b) depict relaxation for two different couplings: $\alpha = 6$ and $9$, with fixed $\beta = \pi/2$ and $J = 1000$, shown on linear and semi-log plots, respectively. The plot illustrates that the relaxation to the saturation is faster when $\alpha$ is more. In (c), we fix $\alpha=6$ and vary $J$ from $500$ to $4000$. The plot suggests that the relaxation time scale increases logarithmically with the dimension of the system. Panel (d) illustrates the corresponding relaxation dynamics on a semi-log plot. Note the horizontal black dashed lines in (a) and (c) along the zero on the $Y$-axis. These lines are drawn to demonstrate that the long-time behavior approaches zero.}
\end{figure}

\subsubsection{Relaxation dynamics}
A sharp relaxation to the saturation follows the initial growth. While the short-time behavior of the OTOCs is well-studied and understood, relaxation is relatively unexplored, with only a few studies. For example, exponential relaxation has been found in random quantum circuit models \cite{bensa2022two} and maximally chaotic Floquet systems \cite{claeys2020maximum}. Relaxation in weakly coupled bipartite systems with chaotic subsystems has been studied in Ref. \cite{prakash2020scrambling}. It has been demonstrated that for fully chaotic systems, the relaxation to the equilibrium follows an exponential scaling \cite{polchinski2015chaos, garcia2018chaos}. 
Given a localized state $|\psi\rangle$ and a fully chaotic $U$, randomization of the state implies that $|\langle\psi|U^t|\psi\rangle|^2\sim e^{-\gamma t}\sim 1/(2J+1)$. Then, the information gets fully scrambled in a time window of $t_{\text{sc}}\sim\log(2J+1)/\gamma$. Note that $t_{\text{sc}}$ can not be smaller than $t_{\text{EF}}$, i.e., $t_{\text{EF}}\lesssim t_{\text{sc}}$. Therefore, in strongly chaotic systems, the relaxation takes place over a window of time $t_{\text{relx}}=(t_{\text{sc}}-t_{\text{EF}})\sim\log(2J+1)$. Beyond this, we expect that the OTOC will get saturated. 
To study the relaxation, we take the quantity $1-C_{F_z=0}(t)/\overline{C_{F_z=0}}$. For $t=0$, we have $C_{F_z=0}(t)=0$, implying that $1-C_{F_z=0}(t)/\overline{C_{F_z=0}}=1$. In the long time limit, this quantity approaches $0$. This is because, as $t\rightarrow\infty$, we have $C_{F_z=0}(t)\sim \overline{C_{F_z=0}}$.
The relaxation dynamics are illustrated in Fig. \ref{largest_relax}. Figure \ref{largest_relax}a and \ref{largest_relax}b demonstrate the relaxation for two different $\alpha$ values while keeping $J=1000$. In Fig. \ref{largest_relax}c and \ref{largest_relax}d, we illustrate the relaxation for different $J$ with $\alpha$ fixed. As time progresses, the curves approach zero with fluctuations. These fluctuations are of the order $10^{-3}$ to $10^{-4}$. We further observe that the relaxation follows an exponential decay for the parameters considered. This result is consistent with the results obtained in Ref. \cite{garcia2018chaos}. 
Though not the central focus of our work, we find the relaxation dynamics an interesting observation subject to further exploration.
\begin{figure}
\includegraphics[scale=0.385]{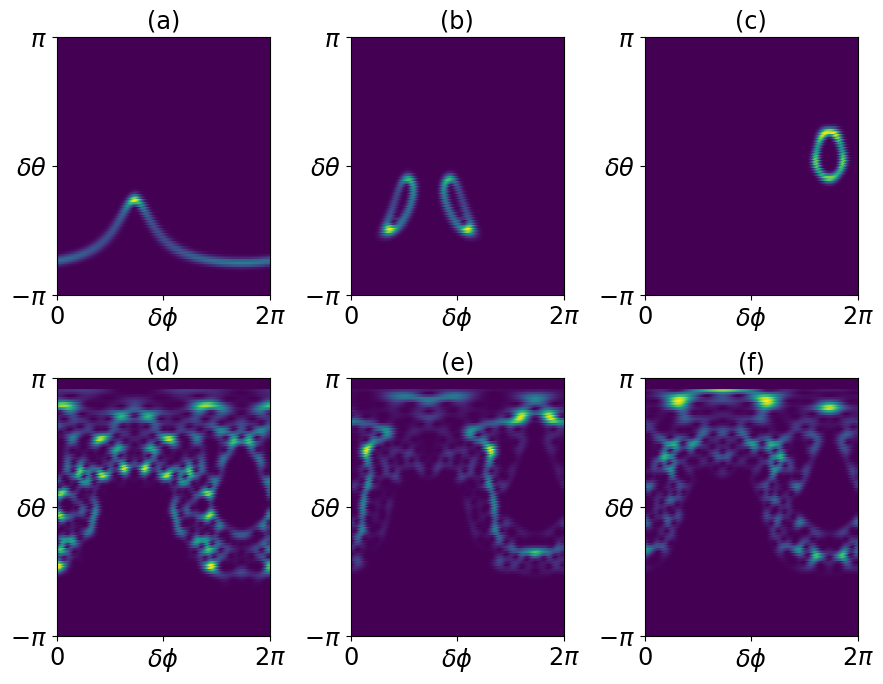}
\caption{\label{fig:eigen_husimi} Husimi plots of the Floquet states of $U_{F_z=0}$. The system parameters correspond to the mixed phase space in the classical limit --- $\alpha=3/2$ and $\beta=\pi/2$. We take $I=J=200$. The Floquet states in (a), (b), and (c) are chosen randomly from the series of points inside the boxes (a), (b), and (c) of Fig. \ref{fig:C_inf_eigen_vs_Sz}, respectively. These three states are localized around different fixed points. (d)-(f) panels correspond to the chaotic Floquet states randomly chosen from the box \ref{fig:C_inf_eigen_vs_Sz}d.  }
\end{figure}

\subsection{Scrambling in the mixed phase space $\left(\alpha=\dfrac{3}{2},\thinspace \beta=\dfrac{\pi}{2}\right)$}\label{Scrambling in the mixed phase space}
In fully chaotic systems, the OTOCs exhibit initial exponential growth and subsequent relaxation, followed by saturation. The saturation value can be predicted using an appropriate random matrix ensemble, such as the COE in the case of the KCT model \cite{haake1991quantum}. However, for the systems with mixed phase space classical limit, a direct correspondence with RMT is absent. Moreover, in these systems, the choice of the initial state largely influences the OTOC growth and saturation, a behavior uncommon in fully chaotic cases. To be precise, the initial states with a significant overlap with the coherent states centered near stable fixed points of the phase space limit the degree of operator scrambling. On the contrary, the scrambling is enhanced if the initial state is localized in the chaotic sea. We observe this behavior in both short-term and long-term dynamics of the OTOC.

To begin, we first consider an initial state, which has a non-zero overlap with the coherent states located on the chaotic sea. Percival's conjecture can be used to construct such states \cite{percival1973regular}. The conjecture categorizes Eigenstates or Floquet states of the system into regular (near stable fixed points) and chaotic states (randomly distributed across chaotic regions). To illustrate this in the KCT model, we show the Husimi plots of six randomly chosen Floquet states corresponding to $\alpha=3/2$, $\beta=\pi/2$, and $J=200$ in Fig. \ref{fig:eigen_husimi}. For a given state, the Husimi function is given by $F_H=\langle \delta\theta, \delta\phi |\rho|\delta\theta, \delta\phi\rangle$, where $|\delta\theta, \delta\phi\rangle$ represents the projection of tensor products of spin coherent states $|\theta_I, \phi_I\rangle \otimes |\theta_J, \phi_J\rangle$ onto the largest invariant subspace ($F_z=0$) \cite{trail2008entanglement}:
\begin{align}\label{spin_coherent state}
|\delta\theta, \delta\phi\rangle =\frac{1}{\mathcal{N}}\sum_{m=-J}^{J}\mu^{m}\frac{(2J)!}{(J-m)!(J+m)!}|m,-m\rangle,
\end{align}
where 
\begin{equation*}
\mu=e^{i\delta\phi/2}\left(\dfrac{1+\sin(\delta\theta /2)}{1-\sin(\delta\theta /2)}\right)
\end{equation*} 
and $\mathcal{N}$ denotes the normalizing constant. In Fig. \ref{fig:eigen_husimi}, the top panels (\ref{fig:eigen_husimi}a, \ref{fig:eigen_husimi}a and \ref{fig:eigen_husimi}c) correspond to the Husimi plots of the Floquet states localized near regular islands. On the other hand, the bottom panels (\ref{fig:eigen_husimi}d, \ref{fig:eigen_husimi}e, and \ref{fig:eigen_husimi}f) represent the states delocalized across the chaotic sea. It's important to note that Percival's conjecture is not always accurate, particularly in the deep quantum regime. Nevertheless, quantities such as Husimi entropy of the states can be employed to roughly distinguish the regular states from the chaotic ones \cite{trail2008entanglement}. As chaotic Floquet states are widespread across the chaotic sea, we choose one such state and initialize the system in it. We then compute the OTOC in Eq. (\ref{commutator}) for the largest subspace by fixing the initial operators $A=B=\hat{S}_z$:
\begin{figure}
\includegraphics[scale=0.5]{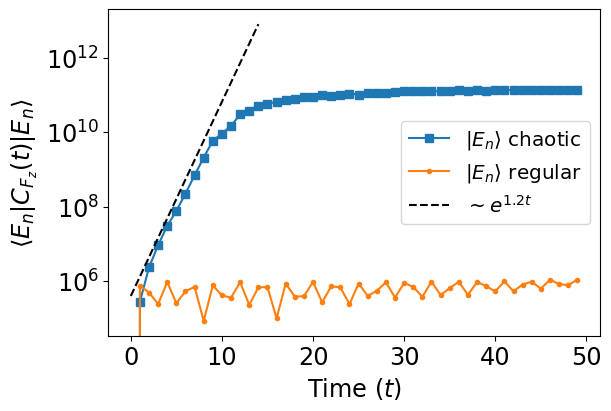}
\caption{\label{fig:OTOC_mixed} Illustration of the OTOC growth in different initial states when the system is associated with the mixed phase space in the $F_z=0$ subsector of the classical phase space. We consider the Floquet states to be the initial states. We fix $\alpha=3/2$, $\beta=\pi/2$ and $J=1000$. The initial operators are $A=B=\hat{S}_z$. In the figure, the blue curve represents the OTOC growth in a chaotic Floquet state. The orange curve corresponds to a regular Floquet state localized near a regular island shown in Fig. \ref{fig:eigen_husimi}a. While the former grows at a rate $\sim e^{1.2 t}$ ($\lambda_{\text{otoc}}\approx 0.6$) initially, the latter one fluctuates around a mean value for all the times. The classical Lyapunov exponent for initial conditions in the chaotic sea is $\lambda_{\text{cl}} \approx 0.5$, determined using Benettin's algorithm. }
\end{figure}
\begin{eqnarray}\label{EN}
\langle E_n|C_{F_z=0}(t)|E_n \rangle= \langle E_n|\left[\hat{S}_z(t), \thinspace \hat{S}_z\right]^{\dagger}\left[\hat{S}_z(t), \thinspace \hat{S}_z\right] |E_n\rangle, 
\end{eqnarray}
where $|E_n\rangle$ denotes a randomly chosen chaotic Floquet state of the system. The corresponding results are shown in Fig. \ref{fig:OTOC_mixed}. In the figure, the blue curve represents the OTOC growth in the chaotic initial state. In this case, the OTOC shows an early time exponential growth ($\sim e^{1.2t}$), followed by saturation. To contrast this with the regular states, we also considered a Floquet state localized near a regular island shown in Fig. \ref{fig:eigen_husimi}a and computed the OTOC in it. In the figure, the regular state OTOC is shown in orange. In this case, the OTOC does not grow and shows fluctuations around its mean value. This clearly indicates that the regular states constrain the scrambling of operators in the mixed-phase space. On the other hand, the chaotic states enhance the scrambling.

\begin{figure}
\includegraphics[scale=0.42]{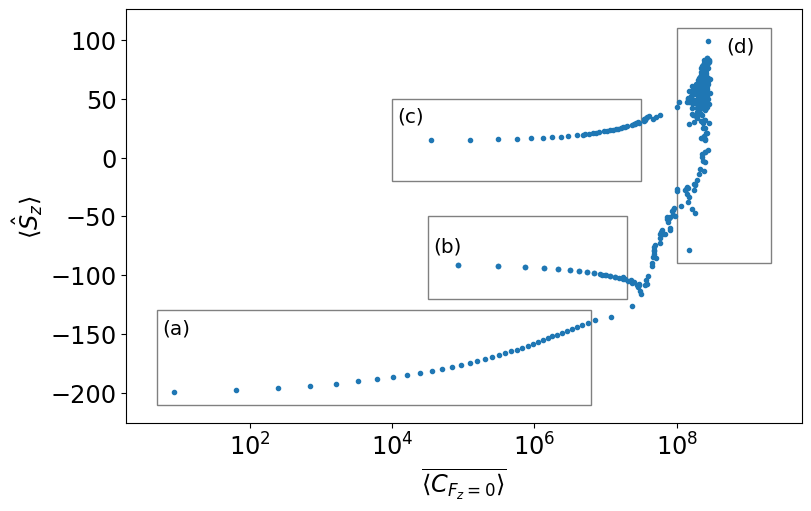}
\caption{\label{fig:C_inf_eigen_vs_Sz} Illustration of $\langle S_z\rangle =\langle E_n| \hat{S}_{z}|E_n\rangle$ versus long-time averaged OTOC with respect the initial state $|E_n\rangle$, when the classical limit of the system is associated with the mixed phase ($\alpha=3/2$ and $\beta=\pi/2$) in the $F_z=0$ sector. The states $\{|E_n\rangle\}$s denote the Floquet states of the operator $U_{F_z=0}$. The angular momentum value is taken to be $I=J=200$. For all the data points, the initial operators are fixed: $A=B=\hat{S}_{z}$. Boxes (a), (b), and (c) correspond to the Floquet states localized on the regular regions of the phase space. Box (d) represents a series of chaotic states that display flat long-time averaged OTOC. Note that the boxes are only representative of the states corresponding to different regions of the phase space and do not characterize the boundaries between these regions. }
\end{figure}

We further elucidate the role of Percival's conjecture in the saturation of mixed-phase space OTOC. To do so, we calculate infinite time averages of the OTOC in the Floquet states and contrast the regular states with the chaotic states:
\begin{eqnarray}\label{floquet_otoc}
\overline{\langle E_n| C_{F_z=0}|E_n\rangle}=\lim_{t\rightarrow\infty}\dfrac{1}{t}\int_{0}^{t}ds\langle E_n|C_{F_z=0}(s)|E_n \rangle,
\end{eqnarray}
where $\langle E_n|C_{F_z=0}(t)|E_n \rangle$ is taken from Eq. (\ref{EN}).
We now cluster the Floquet states based on the infinite time averages and their mean locations in the phase space. 
The mean location can be identified by finding $\langle E_n|\hat{S}_z|E_n\rangle$, which correlates to $\delta\theta$ in the semiclassical limit. Figure \ref{fig:C_inf_eigen_vs_Sz} shows $\overline{\langle E_n|C_{F_z=0}|E_n\rangle}$ vs. $\langle E_n|\hat{S}_z|E_n\rangle$ for a fixed $J=200$ and $\alpha=3/2$. The correlation between $\langle\hat{S}_{z}\rangle$ and $\delta\theta$ causes the Floquet states near stable fixed points to display nearly identical $\langle \hat{S}_z\rangle$ values \cite{gorin1997phase, trail2008entanglement}. In the figure, box (a) corresponds to the series of states localized near the stable south pole [see Fig. \ref{fig:eigen_husimi}a]. The boxes (b) and (c) correspond to the regular islands displayed in Fig. \ref{fig:eigen_husimi}b and Fig. \ref{fig:eigen_husimi}c. Moreover, the chaotic Floquet states typically display higher $\overline{\langle E_n|C_{F_z=0}|E_n\rangle}$ as they are delocalized in the chaotic sea and approximate random quantum states. These states are concentrated vertically along a line within box (d). Note that the plots in Fig. \ref{fig:eigen_husimi}d-\ref{fig:eigen_husimi}f represent three random Floquet states from the box (d). Lesser values of the time-averaged OTOC for the states chosen from boxes (a), (b), and (c) indicate that the operators are less prone to get scrambled if the initial state is localized on a regular island. This analysis shows that the OTOCs are sensitive to the initial state vectors when the classical limit of the system is in a mixed-phase space. Moreover, it is to be noted that the infinite time averages for the regular Floquet states appear to vary smoothly with $\langle \hat{S}_z\rangle$ within the boxes (a), (b), and (c). The points near the extreme right ends of these boxes exhibit infinite time averages close to chaotic states despite being regular. These points represent the states that are localized near the boundaries between the chaotic sea and the regular islands of the phase space. Similar behavior has been previously found for the entanglement entropy \cite{lombardi2011entanglement, madhok2015comment}. Hence, at the boundaries, the operators get scrambled despite the initial states there being regular. Furthermore, our findings complement previous results indicating that hyperbolic fixed points can induce scrambling \cite{hashimoto2020exponential, pilatowsky2020positive, pappalardi2018scrambling, hummel2019reversible, xu2020does, steinhuber2023dynamical}. Conversely, our results demonstrate that regular states located near the boundary of the stable islands with the chaotic region can also induce scrambling.

\begin{figure}
\includegraphics[scale=0.34]{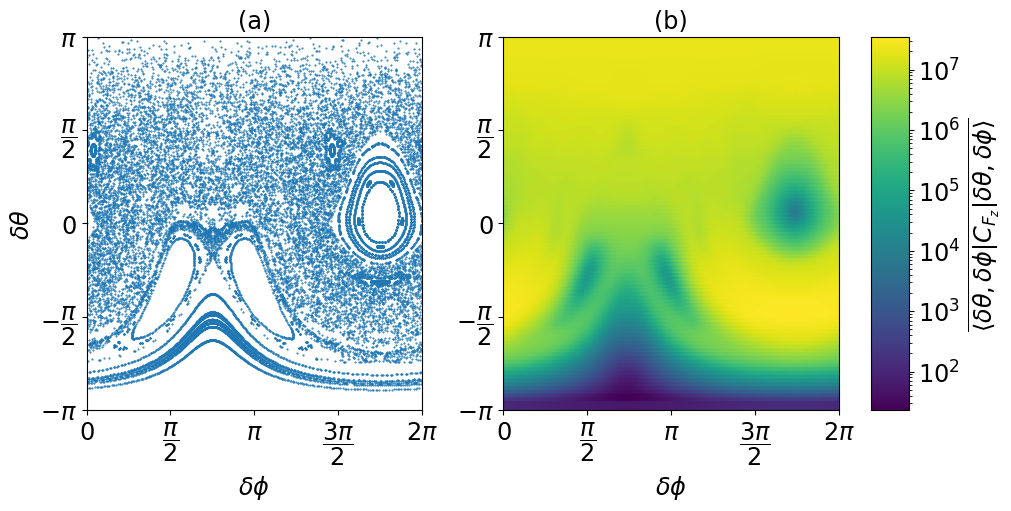}
\caption{\label{fig:husimi_comp} (a) and (b). Side-by-side comparison, showing the infinite-time average of the largest subspace OTOC in the spin coherent states (Fig. (b)) as a superb signature of classical chaos in mixed-phase space ($\alpha =\frac{3}{2}, \beta =\frac{\pi}{2}$) as shown by the Poincaré surface of the section in (a). The angular momentum value is taken to be $I=J=100$. The initial operators are $A=B=\hat{S}_z$. }
\end{figure}

\subsubsection{Quantum-Classical correspondence in the phase space}
Quantum signatures of classical phase space structures appear in various contexts, including the dynamical generation of entanglement \cite{trail2008entanglement, miller1999signatures, ghose2004entanglement, wang2004entanglement, bandyopadhyay2002testing, dogra2019quantum}, tri-partite mutual information \cite{seshadri2018tripartite}, quantum discord \cite{madhok2015signatures} and information gain in quantum state tomography \cite{madhok2014information, madhok2016characterizing}. Do the long-time averages of the OTOCs in the coherent states displays such signatures? To proceed, we replace the Floquet states $\{|E_n\rangle\}$ with the coherent states in the Eq. (\ref{floquet_otoc}) to calculate $\overline{\langle\delta\theta, \delta\phi|C_{F_z=0}|\delta\theta, \delta\phi\rangle}$. Figure \ref{fig:husimi_comp}a and \ref{fig:husimi_comp}b shows the side-by-side comparison between the classical Poincaré section and the scatter plot of the infinite time averaged OTOC (unnormalized) in a mixed-phase space ($\alpha =3/2$). The purpose of this analysis is twofold.
 Firstly, We see a remarkable correlation between the classical phase space and $\overline{\langle\delta\theta, \delta\phi| C_{F_z=0}|\delta\theta, \delta\phi\rangle}$. The latter reproduces classical phase space structures, such as regular islands and the chaotic sea. We see that the values of the quantum calculation attain a fairly uniform value across the entire chaotic sea irrespective of the coordinates of the initial coherent state employed in its computation. Therefore, the information about the initial coordinates of the coherent states in the chaotic sea gets washed away and cannot be recovered from the long-time average OTOC values as seen in the contour plots. This is in contrast with structures seen at the boundary of the chaotic sea and regular islands and also within the regular islands. For example, the darker regions encircle stable fixed points where the operators remain stable and less prone to scrambling. Secondly, while OTOCs have been primarily employed to study the initial rate of divergence of trajectories as captured by Lyapunov exponents, longtime averages of OTOCs provide an understanding of the quantum-classical border.

\begin{table}
\begin{tabular}{ |p{1.5cm}||p{2.1cm}|p{2.1cm}||p{2.cm}|}
\hline
\multicolumn{4}{|c|}{$\alpha=6, \beta=\pi/2$} \\
\hline
Dimension & $\lambda_{F_z=0}$& $\lambda_{\hat{I}_{z}\hat{J}_{z}}$ &$\lambda_{\text{cl}(F_z=0)}$\\
\hline
$J$=500   & 1.7108$\pm O(10^{-4})$ $0\leq t< 4$  &1.47$\pm O(10^{-2})$ $0\leq t< 4$ &\multirow{11}{4em}{$\approx$ 1.55}\\  
\vspace{.15cm} & \vspace{.15cm}&\vspace{.15cm}&\vspace{.15cm}\\
$J$=1000   & 1.7246$\pm O(10^{-4})$ $0\leq t< 4$ & 1.362$\pm O(10^{-2})$ $0\leq t< 4$&\\
\vspace{.15cm} & \vspace{.15cm}&\vspace{.15cm}&\vspace{.15cm}\\
$J$=2000   & 1.6406$\pm O(10^{-3})$ $0\leq t < 5$ & &\\
\vspace{.15cm} & \vspace{.15cm}&\vspace{.15cm}&\vspace{.15cm}\\
$J$=4000   & 1.6988$\pm O(10^{-3})$ $0\leq t< 5$ & &\\
\hline
\end{tabular}
\caption{The table illustrates the classical Lyapunov exponent of the largest subspace ($\lambda_{\text{cl}(F_z=0)}$) and contrasts it with the OTOC growth rate in the largest subspace ($\lambda_{F_z=0}$) for different $J$-values. Further contrast is made between $\lambda_{F_z=0}$ and the growth rate of $C_{\hat{I}_z\hat{J}_z}(t)$, which is denoted by $\lambda_{\hat{I_z}\hat{J}_z}$. The table also provides the time scales over which the exponential fit is considered. We fix the parameters of the system at $\alpha=6$ and $\beta=\pi/2$. Closeness of $\lambda_{F_z=0}$ and $\lambda_{\text{cl}(F_z=0)}$ indicates good quantum-classical correspondence. Moreover, we observe that $\lambda_{\hat{I}_z\hat{J}_z}$ is always slightly less than $\lambda_{F_z=0}$ (see the main text for more details).}
\label{table1}
\end{table}

\begin{figure}
\includegraphics[scale=0.45]{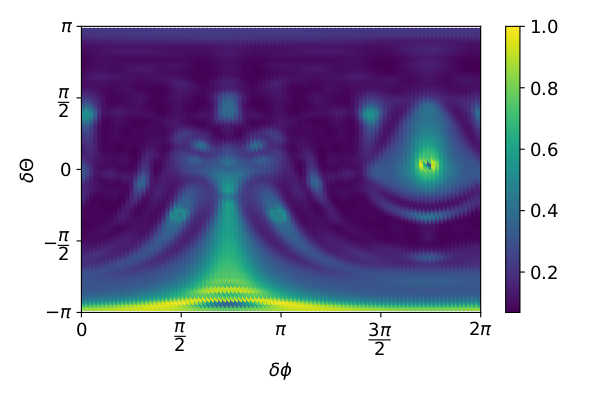}
\caption{\label{fig:otocb} Long time average of the squared commutator when the operators are projectors onto the coherent states. The expectation of the commutator function is taken over a maximally mixed state. The color bar represents the value of the commutator function (normalized).}
\end{figure}

We now consider the initial operators, which are projectors onto the coherent states, i.e., $A=B=|\delta\theta, \delta\phi\rangle\langle\delta\theta, \delta\phi|$. Figure (\ref{fig:otocb}) displays the density plot for the commutator function as a function of the mean coordinate of the coherent states. The plot is reminiscent of the classical Poincar\'e\xspace surface of the section shown in Fig. \ref{fig:husimi_comp}a when $\alpha=3/2$. Intuitively, we can understand the density plot as follows. When the operators are coherent state projectors, the squared commutator becomes 
\begin{equation}
C(t)=|\langle \psi|U(t)| \psi \rangle|^2-|\langle \psi|U(t)| \psi\rangle|^4.
\end{equation}

If the state lies deep inside the regular island close to a stable fixed point, the application of the Floquet operator should not take the state far away from the initial state. This implies that $|\langle \psi|U(t)| \psi \rangle|^2$ is expected to be close to unity. On the other hand, if the state lies in the chaotic sea, then the quantity $|\langle \psi|U(t)| \psi \rangle|^2$ is expected to be close to zero as the application of the Floquet operator immediately randomizes the state. In both cases, the squared commutator will be close to zero, and hence, we observe darker contrast near these regions in the density plot. All the other regions, such as regions in the regular islands away from fixed points and stable orbits, display brighter contrast. In other words, the projectors onto the coherent states in the chaotic sea and the states corresponding to the stable fixed points are less prone to get scrambled.

\subsection{The $\hat{I}_{z}\hat{J}_{z}$ OTOC}\label{the IzJzOTOC}
\begin{figure}
\includegraphics[scale=0.335]{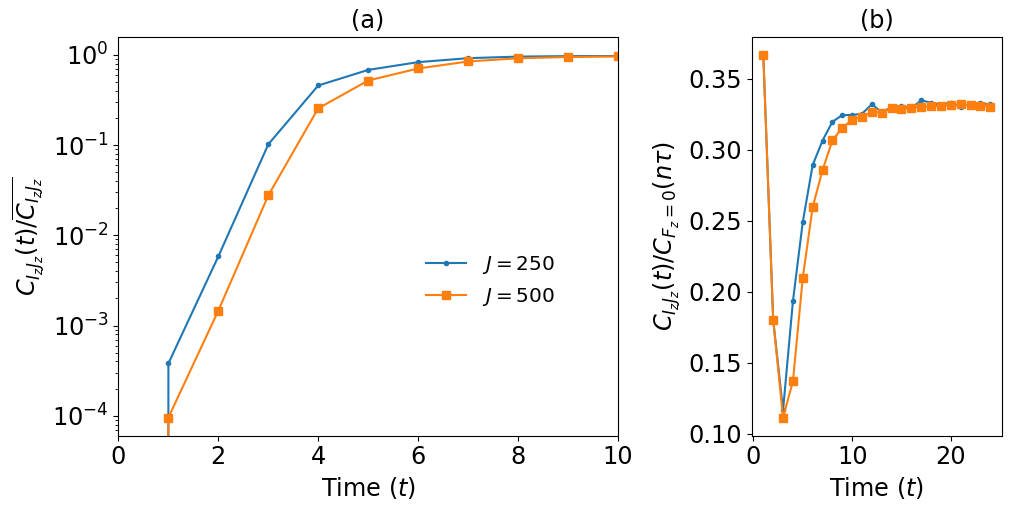}
\caption{\label{fig:zzotoc} (a) Illustration of the early-time exponential growth of $C_{\hat{I}_{z}\hat{J}_{z}}(n\tau)/\overline{C_{\hat{I}_{z}\hat{J}_{z}}}$ for two different $J$ values, where $\overline{C_{\hat{I}_z\hat{J}_{z}}}$ is the long-time average of $C_{\hat{I}_z\hat{J}_z}(n\tau)$. The initial operators are $A=\hat{I}_{z}\otimes \mathbb{I}$ and $B=\mathbb{I}\otimes \hat{J}_{z}$. Here, we fix the parameters $\alpha=6$ and $\beta=\pi/2$. See Table. \ref{table1} for the quantum exponents extracted from the OTOCs. (b) $C_{\hat{I}_{z}\hat{J}_{z}}(n\tau)/C_{F_z=0}(n\tau)$ vs. $n\tau$ for the same $J$ values as before, which aims to elucidate the influence of the largest subspace dynamics on the overall OTOC dynamics. }   
\end{figure}

Here, we compute the OTOC for the entire system of KCT by considering the initial operators $A=\hat{I}_{z}\otimes \mathbb{I}$ and $B=\mathbb{I}\otimes \hat{J}_{z}$. Since $A$ and $B$ commute with $F_z$, they can be decomposed as $A=\oplus_{F_z}A_{F_z}$, and $B=\oplus_{F_z}B_{F_z}$. Accordingly, the commutator function $C(t)$ becomes
\begin{equation}\label{F_z_conserving}
C_{\hat{I}_{z}\hat{J}_{z}}(t)=\dfrac{1}{(2J+1)^2} \sum_{F_z}\Tr[A_{F_z}^2(t)B_{F_z}^2]-\Tr[A_{F_z}(t)B_{F_z}A_{F_z}(t)B_{F_z}]
\end{equation}
where $A_{F_z}(t)=U^{\dagger t}_{F_z}A_{F_z}U^{t}_{F_z}$. Given the absence of cross-terms between different subspaces, each subspace makes an independent contribution to the OTOC. We shall see from the numerical results that $C_{\hat{I}_{z}\hat{J}_{z}}(t)$ shows a growth rate slightly less than that of $C_{F_z=0}(t)$ over a short period. This can be intuitively understood by examining the following quantity:
\begin{equation}
\dfrac{C_{\hat{I}_{z}\hat{J}_{z}}(t)}{C_{F_z=0}(t)}=\dfrac{1}{2J+1}\left[1+\sum_{\substack{F_z=-2J\\ F_z\neq 0}}^{2J}\left(1-\dfrac{|F_z|}{2J+1}\right)\dfrac{C_{F_z}(t)}{C_{F_z=0}(t)}\right], 
\end{equation}
where $C_{F_z}(t)$ denotes OTOC contribution from a subspace with quantum number $F_z$. Due to the fully chaotic nature, the largest subspace will likely have the shortest Ehrenfest time. For $t\lesssim t_{\text{EF} (F_z=0)}$, there is a high chance that in the subspaces close to the largest one, the OTOCs will grow exponentially with the rates $\lambda_{F_z\neq 0}\lessapprox\lambda_{F_z=0}$. Consequently, $C_{F_z\neq 0}(t)/C_{F_z=0}(t)\sim\exp{2(\lambda_{F_z\neq 0}-\lambda_{F_z=0})t}$ either decays exponentially but slowly or stays a constant. On the other hand, for $|F_z|\gg 0$, the quantity $C_{F_z\neq 0}(t)/C_{F_z=0}(t)$ decays quickly to zero. Therefore, $C_{\hat{I}{z}\hat{J}{z}}(t)/C_{F_z=0}(t)$ most likely decays for $t\lesssim t_{\text{EF}(F_z=0)}$. As a result, the initial growth rate of $C_{\hat{I}{z}\hat{J}{z}}(t)$ is expected to be slightly lower than that of $C_{F_z=0}(t)$. We confirm this from the numerical simulations shown in Fig. \ref{fig:zzotoc}. For the numerical simulations, we consider $\alpha=6$, which is strong enough to make many subspaces fully chaotic. 

In Fig. \ref{fig:zzotoc}a, we plot the short-time exponential growth of the OTOC for two different $J$ values. The plot is normalized by dividing $C_{\hat{I}_{z}\hat{J}_{z}}(t)$ with its infinite time average. Table. \ref{table1} makes a comparison between $\lambda_{\hat{I}_{z}\hat{J}_{z}}$ and $\lambda_{F_z=0}$, and we see that the former is always slightly less than the latter, as inferred earlier. The quantity $C_{\hat{I}_{z}\hat{J}_{z}}(t)/C_{F_z=0}(t)$ is examined for $t\geq 1$ in Fig. \ref{fig:zzotoc}b. The ratio initially decays with time for $1\leq t\leq 3$, indicating that the largest subspace dominates. The decay time scale correlates with $t_{\text{EF} (F_{z}=0)}$. The ratio will grow afterward until the saturation, suggesting non-trivial contributions from the other subspaces.

\begin{figure}
\includegraphics[scale=0.335]{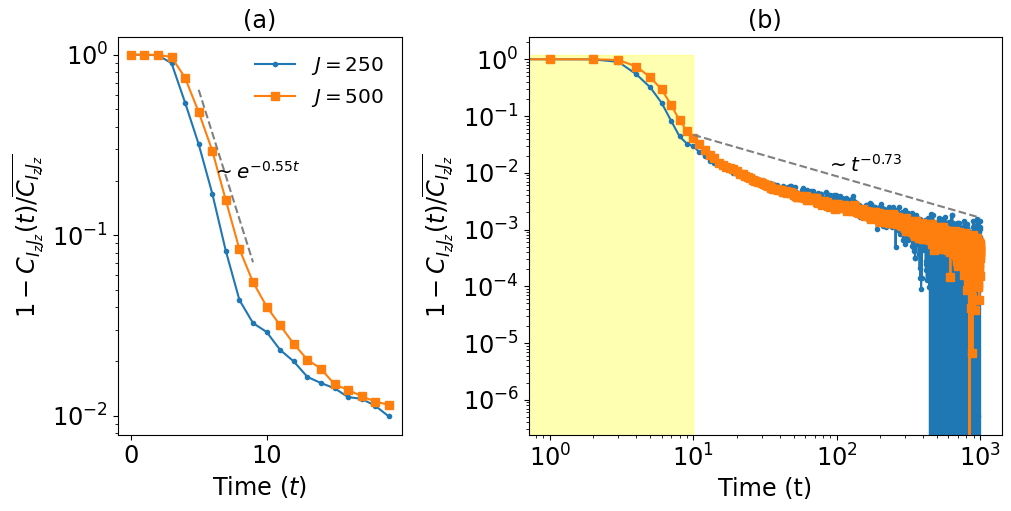}
\caption{\label{zz_relax} Illustration of the two step relaxation of the $I_z-J_z$ OTOC. The parameters are fixed at $\alpha=6$ and $\beta=\pi/2$. The results are shown for the same $J$ values considered in Fig. \ref{fig:zzotoc}. The relaxation appears to proceed in two steps. Panel (a) depicts the short-time exponential decay of $1-C_{\hat{I}_{z}\hat{J}_{z}}(n\tau)/\overline{C_{\hat{I}_{z}\hat{J}_{z}}}$ during the first stage of relaxation. The largest subspace dynamics largely dominate this phase. For $J=500$, we observe a decay that scales as $\sim e^{-0.55 t}$. (b). Long-time algebraic decay of $1-C_{\hat{I}_{z}\hat{J}_{z}}(n\tau)/\overline{C_{\hat{I}_{z}\hat{J}_{z}}}$ in the second-phase.}   
\end{figure}

We now study the relaxation dynamics using the quantity $1-C_{\hat{I}_{z}\hat{J}_{z}}(t)/\overline{C_{\hat{I}_{z}\hat{J}_{z}}}$. Recall that the subspace dynamics transition from chaotic to regular as $|F_z|$ is increased from zero. Consequently, we observe (i) longer Ehrenfest times with slower growth rates and (ii) lesser saturation values as predicted by RMT in those subspaces. Due to the observations (i) and (ii), $C_{\hat{I}_{z}\hat{J}_{z}}(t)$ takes a longer time to saturate compared to $C_{F_z=0}(t)$. The numerical results are shown in Fig. \ref{zz_relax}. The relaxation, as the figure suggests, proceeds in two steps. The first phase (see Fig. \ref{zz_relax}a), which is considerably dominated by the largest subspace dynamics, displays an exponential decay. The decay exponent correlates with that of $1-C_{F_z=0}(t)/\overline{C_{F_z=0}}$. For $J=500$, the decay exponent is obtained to be $\approx 0.55$. At the end of this phase, the OTOC contributions from the subspaces close to $F_z=0$ almost reach a saturation point. The relaxation then shows a crossover to an algebraic law $\sim t^{-0.73}$, where the smaller subspace dynamics dominate. Moreover, due to the regular dynamics of those subspaces, we observe large fluctuations in the quantity $1-C_{\hat{I}_{z}\hat{J}_{z}}(t)/\overline{C_{\hat{I}_{z}\hat{J}_{z}}}$ as $t$ approaches infinity. The corresponding numerics are shown in Fig. \ref{zz_relax}b. As $\alpha$ increases further, smaller subspaces eventually become chaotic, leading to sharper decay in the first phase and flatter dynamics in the second phase.

\subsection{The $\hat{I}_{x}\hat{J}_{x}$ OTOC}\label{the IxJxOTOC}
Here, we study the operator growth for the local operators that do not commute with $\hat{F}_z$. Specifically, we take $A=\hat{I}_x\otimes \mathbb{I}$ and $B=\mathbb{I}\otimes \hat{J}_x$. Unlike $\hat{I}_z$ and $\hat{J}_z$, these operators do not admit the decomposition into disjoint invariant subspaces. Thus, OTOC dynamics are mixed across various subspaces. We aim to contrast the scrambling in this case with the previous one. First, for the initial operators $A$ and $B$, it is useful to note that the following:
\begin{eqnarray}\label{inter}
\left. 
  \begin{array}{ c l }
U_{F_z}^{\dagger}AU_{F'_z}\neq\mathbf{0}\\
U_{F_z}^{\dagger}BU_{F'_z}\neq\mathbf{0}
\end{array}
\right\}\quad\text{iff}\quad F'_z=F_z\pm 1,
\end{eqnarray}
where $U_{F_z}$ remains $(2J+1)^2$-dimensional, but acts non-trivially on the $F_z$-subspace alone. For more details concerning Eq. (\ref{inter}), refer to Appendix \ref{Appa}. We reemphasize that the $F_z=0$ subspace typically drives the overall OTOC dynamics. However, the observation in Eq. (\ref{inter}) results in vanishing $4$-point correlators at all the subspace levels, i.e., 
\begin{equation}
\Tr\left[U^{\dagger t}_{F_z}AU^{t}_{F_z}BU^{\dagger t}_{F_z}AU^{t}_{F_z}B\right]=0\quad \forall F_{z} \text{ and } t\geq 0.
\end{equation}
On the contrary, the $4$-point correlators that connect adjacent subspaces as given by $\Tr[U^{\dagger t}_{F_z}AU^{t}_{F_z\pm 1}BU^{\dagger t}_{F_z}AU^{t}_{F_z\pm 1}B]$ remain non-vanishing. The intertwining of adjacent subspace evolutions $U_{F_z}$ and $U_{F_z\pm 1}$ causes a slow down of the OTOC dynamics whenever the largest subspace is supposedly dominant.

\begin{figure}
\includegraphics[scale=0.335]{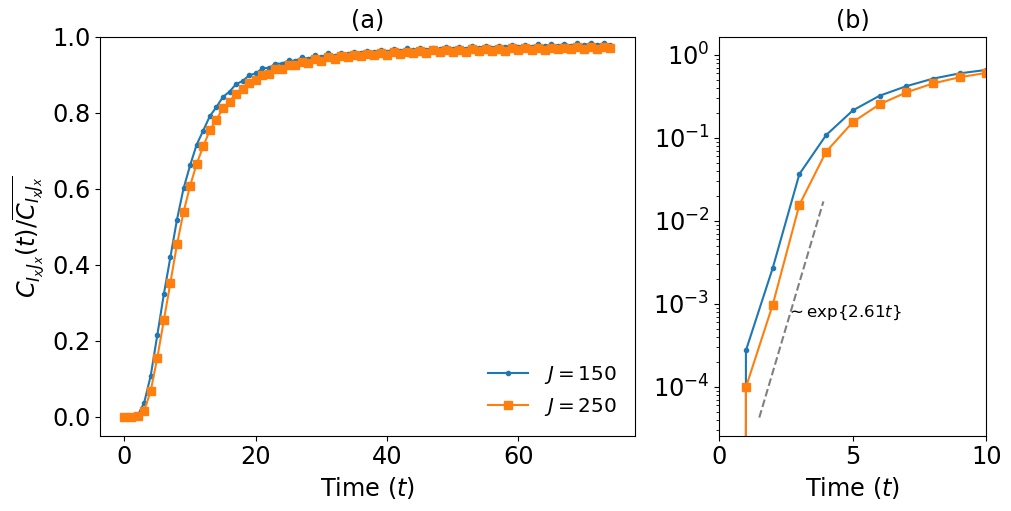}
\caption{\label{xxotoc} (a) The OTOC when the initial operators are $A=\hat{I}_{x}\otimes \mathbb{I}$ and $B=\mathbb{I}\otimes \hat{J}_{x}$ for two different $J$ values. The parameters are kept fixed at $\alpha=6$ and $\beta=\pi/2$. (b) Illustration of the early-time exponential growth on the semi-log scale. The OTOC for $J=250$ grows initially at a rate $\sim e^{2.61 t}$.}   
\end{figure}

We now numerically calculate $C_{\hat{I}_{x}\hat{J}_{x}}(t)$ for two different values of $J$, namely, $J=150$ and $J=250$. The corresponding results are plotted in Fig. \ref{xxotoc}. Figure \ref{xxotoc}a demonstrates growth of $C_{\hat{I}_{x}\hat{J}_{x}}(t)/\overline{C_{\hat{I}x\hat{J}_x}}$ and the corresponding short-time exponential growth is depicted in Fig. \ref{xxotoc}b. The parameters $\alpha$ and $\beta$ are kept fixed at $6$ and $\pi/2$, respectively, as before. The exponential growth appears within the regime $1\leq t\leq 3$ for both values of $J$. Through the exponential fitting, we find $\lambda_{\hat{I}_{x}\hat{J}_{x} (J=150)}\approx 1.305$ and $\lambda_{\hat{I}_{x}\hat{J}_{x}(J=250)}\approx 1.388$, which are roughly about the same as those of $\lambda_{\hat{I}_{z}\hat{J}_{z}}$, but with a minute difference (see Table. \ref{table1}). However, the OTOC displays slower relaxation in the intermediate regime than in the previous case. For instance, for $J=250$, we observe $(1-C_{\hat{I}_{x}\hat{J}_{x}}(t)/\overline{C_{\hat{I}_{x}\hat{J}_{x}}})\sim \exp\{-0.15 t\}$, which is in contrast with the previous case, where $(1-C_{\hat{I}_{z}\hat{J}_{z}}(t)/\overline{C_{\hat{I}_{z}\hat{J}_{z}}})\sim \exp\{-0.6 t\}$. Hence, the above inference about the slowdown of OTOC dynamics is justified. 
 
\begin{figure}
\includegraphics[scale=0.335]{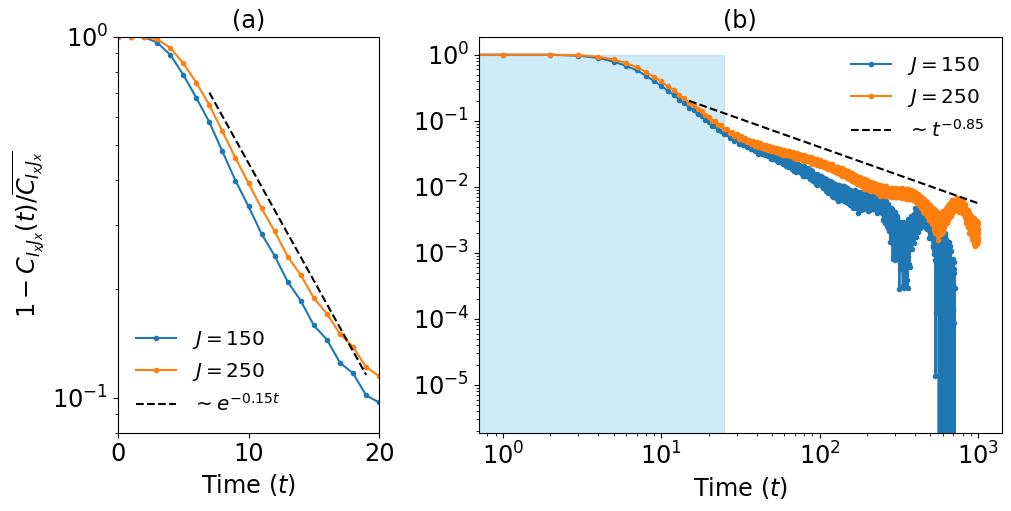}
\caption{\label{xxotoc_relax} Relaxation dynamics of the $\hat{I}_x-\hat{J}_x$ OTOC. We observe that the relaxation, similar to $\hat{I}_z-\hat{J}_z$, proceeds in two steps. In (a), the short-time exponential decay of $1-C_{\hat{I}_{x}\hat{J}_{x}}(n\tau)/\overline{C_{\hat{I}_{x}\hat{J}_{x}}}$ is shown, which is slower than its $I_z-J_z$ counterpart (see Fig. \ref{zz_relax}a). (b). Long-time relaxation following algebraic decay $\sim t^{-1.065}$ and $\sim t^{-0.85}$ respectively for $J=150$ and $J=250$.}   
\end{figure}

Beyond the exponential decay, the quantity $1-C_{\hat{I}_x\hat{J}_x}(t)/\overline{C_{\hat{I}_x\hat{J}_x}}$ displays a crossover to an algebraic law. The results are shown in Fig. \ref{xxotoc_relax}. For J=250, through the power law fitting of the data in this regime, we find that the relaxation follows an approximate scaling $\sim t^{-0.85}$, which is slightly faster than the earlier case of $1-C_{\hat{I}_{z}\hat{J}_{z}}(t)/\overline{C_{\hat{I}_{z}\hat{J}_{z}}}\sim t^{-0.75}$. Moreover, due to the regular dynamics associated with the smaller subspaces of the KCT model, the saturation is usually accompanied by fluctuations around the mean value of the OTOC.

\subsubsection{RMT prediction for the saturation of $\hat{I}_x\hat{J}_{x}$ OTOC}
The OTOCs in fully chaotic systems are expected to saturate after a sufficiently long time. The saturation value follows if we replace the Floquet operator $U^{t}$ with a set of random unitaries drawn from the appropriate ensemble \cite{lakshminarayan2019out}. Recall that the standard COE ensemble is the appropriate random matrix ensemble for the kicked coupled top. This is then followed by performing an average over the COE unitaries.
We denote the resultant OTOC with $C_{\text{RMT}}$ for the given operators $A$ and $B$. Moreover, to incorporate the symmetry operator $\hat{F}_z$, we generate the random COE unitaries of the form $U=\oplus_{F_z}U_{F_z}$, where $U_{F_z}\in \text{COE}(2J+1-|F_z|)$. The observation in Eq. (\ref{inter}) assists in simplifying the COE average \cite{brouwer1996diagrammatic} of the two-point and four-point correlators as follows:
\begin{eqnarray}\label{23}
\overline{C_{2}}&=&\dfrac{1}{(2J+1)^2}\sum_{F_z}\Tr\left[\overline{U^{\dagger}_{F_z}A^2U_{F_z}B^2}\right]\nonumber\\
&=&\dfrac{1}{(2J+1)^2}\sum_{F_z}\dfrac{\Tr([A^2]_{F_z}^TB^2)+\Tr([A^2]_{F_z})\Tr([B^2]_{F_z})}{2J+2-|F_z|}\nonumber\\
\end{eqnarray}
and 
\begin{eqnarray}\label{24}
\overline{C_{4}}&=&\dfrac{1}{(2J+1)^2}\sum_{F_z}\Tr\left[\overline{U^{\dagger}_{F_z}AU_{F_z\pm 1}BU^{\dagger}_{F_z}AU_{F_z\pm 1}B}\right]   \nonumber\\ 
&=& 0,
\end{eqnarray}
where $[A^2]_{F_z}=\mathbb{I}_{F_z}A^2\mathbb{I}_{F_z}$ and $I_{F_z}$ is an identity operator acting exclusively on  the $F_z$-subspace, i.e., $\mathbb{I}_{F_z}\equiv\mathbb{I}_{F_z}\oplus\mathbb{O}_{\text{rest}}$, the null matrix $\mathbb{O}_{\text{rest}}$ acts on rest of the subspaces. For a detailed derivation of Eqs. (\ref{23}) and (\ref{24}), refer to Appendix \ref{AppB}. It then follows that $C_{\text{RMT}}=\overline{C_2}$.

\section{Scrambling, operator entanglement and coherence}\label{Scrambling, operator entanglement and coherence}
The preceding section focused on the analysis of OTOCs in the KCT system for various choices of initial operators. In this section, we present a general result concerning the OTOCs and the operator entanglement of arbitrary quantum evolutions. The results in this section are independent of those presented in the previous section. Recent studies have demonstrated that in a bipartite quantum system, when initial operators are local, independent, and uniformly chosen at random from the unitary group, the averaged OTOC equates to the linear operator entanglement entropy of the corresponding quantum evolution \cite{styliaris2021information, pawan}. This result primarily relies on selecting Haar-random local unitaries as initial operators. However, it remains unclear whether such a relationship persists when the initial operators are different from Haar random unitaries. Here, we address this question by considering random Hermitian operators drawn from the unitarily invariant Gaussian ensemble (GUE) as the initial operators. We first establish that the connection between the OTOC and the operator entanglement holds up to a scaling factor for the GUE operators. Additionally, we examine diagonal GUE matrices as initial operators and demonstrate their relationship with the coherence-generating power introduced in Ref. \cite{anand2021quantum}.

\subsection{GUE initial operators}\label{GUE operators}   
We first consider a bipartite quantum system governed by the evolution $U$. We label the subsystems with $A$ and $B$ and assume that both are equal dimensional, i.e., $d_1=d_2=d$. We denote with $C_{PQ}(t)$, the OTOC for two initial operators $P=O_1\otimes\mathbb{I}$ and $Q=\mathbb{I}\otimes O_2$, where $O_1, \thinspace O_2\in\text{GUE}(d)$. We are interested in evaluating the quantity $C_{\text{GUE}}(t)=\overline{C_{PQ}}(t)=\overline{C_{2}}(t)-\overline{C_{4}}(t)$, where the overline indicates the averaging over all $O_1$ and $O_2\in \text{GUE}(d)$. In the following, we first evaluate $\overline{C_{2}}(t)$:
\begin{eqnarray}
\overline{C_{2}}(t)&=&\dfrac{1}{d^2}\int_{O_1\in\text{GUE}}d\mu(O_1)\int_{O_2\in\text{GUE}}d\mu(O_2)\Tr\left( U^{\dagger t}O^2_1U^{t}O^{2}_{2} \right)\nonumber\\
&=&\dfrac{1}{d^2}\Tr\left[U^{\dagger t}\left(\int_{O_1} d\mu(O_1)O^{2}_1\right)U^{t}\left(\int_{O_2} d\mu(O_2)O^{2}_{2}\right)\right]
\end{eqnarray}
where $d\mu(O_{1(2)})$ represents the normalized measure over GUE($d$). Then, the following integral follows from the unitary invariance property of the GUE measure. 
\begin{eqnarray}\label{osqre}
\int_{O_1\in \text{GUE}} O^{2}_{1} d\mu(O_1)&=&\int_{O_1\in \text{GUE}} d\mu(O_1)\int_{u\in U(d)} d\mu(u) u^{\dagger}O^{2}_{1}u,\nonumber\\ 
&=&\int_{O_1\in\text{GUE}}d\mu(O_1)\left(\dfrac{\Tr(O^{2}_{1})}{d}\right)\mathbb{I}_{d}\nonumber\\
&=&d\left(\mathbb{I}_{d}\right),
\end{eqnarray}
where $d\mu(u)$ is the invariant Haar measure over the unitary group $U(d)$. 
It then follows that
\begin{eqnarray}
\overline{C_2}(t)=\dfrac{1}{d^2}\Tr(U^{\dagger t}U^t)=d^2
\end{eqnarray}

We now evaluate the average four-point correlator.  
\begin{eqnarray}\label{4poin}
\overline{C_4}(t)&=&\dfrac{1}{d^2}\int_{O_1}d\mu(O_1)\int_{O_2}d\mu(O_2)\Tr\left(O_1(t)O_2O_1(t)O_2\right)\nonumber\\
&=&\dfrac{1}{d^2}\int_{O_1}d\mu(O_1)\int_{O_2}d\mu(O_2)\Tr(S U^{\dagger t\otimes 2}O^{\otimes 2}_1U^{t\otimes 2}O_2^{\otimes 2}),\nonumber\\ 
\end{eqnarray}
where, in the second equality, a replica trick given in Ref. \cite{styliaris2021information} is used --- for any $d$-dimensional $p$ and $q$ matrices, the replica trick implies $\Tr(pq)=\Tr(S(p\otimes q))$. Also note that $S$ in Eq. (\ref{4poin}) swaps quantum states on the Hilbert space $\mathcal{H}_{AB}\otimes \mathcal{H}_{AB}$. This allows us to write $S$ as $S=S_{AA}S_{BB}$. To evaluate $\overline{C_{4}}(t)$, we again invoke the unitary invariance property of the probability measure of GUE.
\begin{equation}
\int_{O_1}d\mu(O_1) O^{\otimes 2}_{1} =\int_{O_1}d\mu(O_1)\int_{u\in U(d)} d\mu(u) \left(u^{\dagger \otimes2}O^{\otimes 2}_1u^{\otimes 2}\right),
\end{equation}
where the second integral on the right-hand side can be solved using the Schur-Weyl duality \cite{zhang2014matrix}. 
\begin{eqnarray}
\int_{u\in U(d)} d\mu(u) \left(u^{\dagger \otimes2}O^{\otimes 2}_{1}u^{\otimes 2}\right)&=&\left[ \left( \frac{\Tr(O_1)^2}{d^2-1}-\frac{\Tr(O^{2}_1)}{d(d^2-1)} \right)\mathbb{I}_{d^2}\right.\nonumber\\
&&\left.- \left(\frac{\Tr(O_1)^2}{d(d^2-1)}-\frac{\Tr(O^{2}_1)}{d^2-1}\right)S_{AA} \right].\nonumber\\
\end{eqnarray}
We are now left with the integrals of $\Tr(O_1)^2$ and $\Tr(O^{2}_{1})$ over GUE, which are related to the mean and variance of the elements of the GUE matrices. Upon performing these integrals, we obtain 
\begin{eqnarray}\label{31}
\int_{O_1\in \text{GUE}}d\mu(O_1) O^{\otimes 2}_{1}  = S_{AA}. 
\end{eqnarray}
Similarly, 
\begin{eqnarray}\label{32}
\int_{O_2\in \text{GUE}}d\mu(O_2) O^{\otimes 2}_{2}  = S_{BB}. 
\end{eqnarray}
As a result, the four-point correlator averaged over GUE local operators can be written as 
\begin{eqnarray}
\overline{C_4}(t)&=&\dfrac{1}{d^2}\Tr[S (U^{\dagger t})^{\otimes 2}S_{AA}(U^{t})^{\otimes 2}S_{BB}]    \nonumber\\
&=& \dfrac{1}{d^2}\Tr\left( S_{AA}(U^{\dagger t})^{\otimes 2} S_{AA}(U^{\dagger t})^{\otimes 2}\right)
\end{eqnarray}
It then follows that  
\begin{eqnarray}
C_{\text{GUE}}(t)=d^2-\frac{1}{d^2}\Tr[S_{AA} (U^{\dagger t})^{\otimes 2}S_{AA}(U^{t})^{\otimes 2}], 
\end{eqnarray}
where the second term is related to the linear entanglement entropy of quantum evolutions \cite{styliaris2021information}. In deriving the above relation, we utilized the unitary invariance of the measure over GUE. Additionally, the mean ($\mu=0$) and variance ($\sigma^2=1$) of the elements of the GUE operators were crucial for deriving this relation. We note that the unitary invariance is sufficient to establish a relation between the operator entanglement and the OTOC. On the other hand, zero mean, and unit variance ensure that Eqs. (\ref{31}) and (\ref{32}) hold. Therefore, as long as the operators are chosen from ensembles with the listed properties, such as unitary invariance, zero mean and unit variance, the above relation is expected to remain valid.

\subsection{Random diagonal initial operators}\label{Random diagonal initial operators}
Here, we take $O_1$ and $O_2$ to be diagonal operators with the elements chosen randomly from the Gaussian distribution with the mean $\mu=0$ and the standard deviation $\sigma=1$, i.e., 
\begin{eqnarray}
O_1=\sum_{i=0}^{d-1}x_i|i\rangle\langle i|,\quad \text{and}\quad O_2=\sum_{i=0}^{d-1}y_i|i\rangle\langle i|,  
\end{eqnarray}
where $x_i$ and $y_i$ are independent Gaussian random variables. It then follows that
\begin{eqnarray}
\int_{O_{1(2)}}O^2_{1(2)}d\mu(O_{1(2)})&=&\mathbb{I}_{1(2)}, \nonumber\\
\text{and}\quad\int_{O_{1(2)}}(O_{1(2)}\otimes O_{1(2)})d\mu(O_{1(2)})&=&\sum_{i, j=0}^{d-1}\delta_{ij}|ij\rangle\langle ij|. 
\end{eqnarray}
The two-point and the four-point correlators, in this case, take the following forms:
\begin{eqnarray}
\overline{C_2}(t)&=& d^2\nonumber\\
\overline{C_4}(t)&=&\sum_{ijkl}\langle ji|U^{\dagger t\otimes 2}|kk\rangle\langle kk|U^{t\otimes 2}|ll\rangle\delta_{li}\delta_{lj}\nonumber\\
&=&\sum_{kl}\langle kk|U^{t\otimes 2}(t)|ll\rangle\langle ll|U^{\dagger t\otimes 2}|kk\rangle\nonumber\\
&=&\sum_{kl}|\langle k|U^t|l\rangle|^4
\end{eqnarray}
Hence, the commutator function is given by
\begin{eqnarray}
C_{\text{dGUE}}(t)=1-\dfrac{1}{d^2}\sum_{kl}|\langle k|U^t|l\rangle|^4.
\end{eqnarray}
This is closely related to the coherence generating power (CGP) of the time evolution operator \cite{anand2021quantum}.

\section{SUMMARY AND DISCUSSIONS}\label{SUMMARY AND DISCUSSIONS}
The central question concerning quantum chaos is how classically chaotic dynamics inform us about specific properties of quantum systems, e.g., the energy spectrum, nature of eigenstates, correlation functions, and, more recently, entanglement. Alternatively, what features of quantum systems arise since their classical description is chaotic? It is now well understood that if the quantum system has an obvious classical limit, then the dynamics of quantities like entanglement entropy and the OTOC are characterized by the corresponding classical Lyapunov exponents (see for example \cite{lerose2020bridging} and references therein). Recently, the study of OTOC and scrambling of quantum information as quantified by operator growth has witnessed a surge in interest. This is supplemented by the role of non-integrability, symmetries, and chaos in the above and in the study of thermalization and statistical aspects of many-body quantum systems. Our work attempts to explore operator scrambling and OTOCs and the role of symmetries and conserved quantities in the process.
For this purpose, we have employed a model system of KCTs that is simple enough yet exhibits rich dynamics.
Many-body quantum systems have an intimate connection with quantum chaos. There has been a significant push to understand the issues involving thermalization, irreversibility, equilibration, coherent backscattering, and the effects of quantum interference in such systems. The Eigenstate thermalization hypothesis (ETH) offers a rich platform for understanding several of these phenomena. However, one needs to be extra careful in attributing chaos and non-integrability to ETH, as recent studies have shown that even a classical Lipkin-Meshkov-Glick (LMG) model model displays ETH properties \cite{kelly2020thermalization, lambert2021quantum}. In our work, we have considered the simplest many-body system, the kicked coupled top, which is rich enough to allow us to study these aspects. We have taken up this line of study and explored how information propagates with operator scrambling and how the incompatibility of two operators can quantify this with time.

We have numerically studied the OTOCs in the KCT model, which exhibits chaos in the classical limit and conserves the angular momentum along the $z$-axis. The conservation law enables us to partition the system into distinct invariant subspaces with constant magnetization. We first considered the largest invariant subspace and examined the scrambling in the fully chaotic regime. We observed a good quantum-classical correspondence between the OTOC growth rate and the maximum classical Lyapunov exponent. We then studied the mixed phase space OTOC dynamics with the help of Percival's conjecture. We have observed that the operators are more likely to get scrambled if the initial state is supported over the chaotic sea. This is reflected both in the short-time as well as the long-time behavior. Additionally, we have studied the infinite time averages of the OTOCs in the Floquet states against their mean location in the phase space. We observed that states corresponding to the same regular islands form clusters. Interestingly, these averages appear to vary smoothly with respect to the mean location of these states. Conversely, the OTOC time averages in chaotic Floquet states take nearly uniform values. Also, the Floquet states that are localized near the boundaries exhibit infinite time averages close to chaotic states despite being regular. It is to be noted that similar results exist for the entanglement entropy \cite{lombardi2011entanglement, madhok2015comment}. Our findings complement previous results indicating that hyperbolic fixed points can induce scrambling without chaos \cite{hashimoto2020exponential, pilatowsky2020positive, pappalardi2018scrambling, hummel2019reversible, xu2020does, steinhuber2023dynamical}. In particular, our results demonstrate that regular states located near the boundary of the stable islands with the chaotic region can also induce scrambling. We have also studied the OTOCs by taking coherent states as initial states. Interestingly, while the short-time growth of OTOCs captures the Lyapunov divergence, the long-term behavior of OTOCs remarkably reproduces classical phase space structures.

Subsequently, we extended our analysis to the entire KCT system, investigating OTOCs for two distinct choices of initial operators. We found that the scrambling behavior differs depending on the choice of initial operators. Furthermore, conservation laws have been observed to impede operator dynamics, regardless of the choice of the initial operators. We also proved an independent result concerning the OTOCs and operator entanglement entropy of arbitrary time evolution operators by taking the Gaussian unitary operators as the initial operators \cite{styliaris2021information}. Moreover, the OTOC is related to the coherence generating power for the diagonal Gaussian operators \cite{anand2021quantum}. It is worth noting that in this work, we have considered equal spin sizes for the OTOC analysis. Hence, it is interesting to see if considering unequal magnitudes results in different dynamics \cite{haake1991quantum}. Nevertheless, we do not expect significant differences as the time-reversal symmetry is preserved in the system. There has been debate whether or not ``scrambling" is necessary or sufficient for chaos \cite{dowling2023scrambling, xu2020does} as OTOCs can show rapid exponential growth from unstable fixed points. We feel one needs both short time and long time behavior to capture the complete dynamics of the system. Not only does this give a more complete picture when using OTOCs as probes, but it also settles the question that OTOCs can be used as probes for chaos when studied in both limits.

One of the central focuses in quantum information science is to simulate these systems on a quantum device and on the related issues involving quantifying the complexity of these simulations, benchmarking these simulations in the presence of errors, and exploring connections to quantum chaos in these systems. How is information scrambling related to the complexity of simulating many-body quantum systems \cite{sieberer2019digital}? We hope our study paves the way for the exploration of these intriguing directions.

\begin{acknowledgments}
We are grateful to Arul Lakshminarayan for useful discussions.
This work was supported in part by grant number SRG/2019/001094/PMS from SERB and MHRD/DST grants SB20210807PHMHRD008128, SB20210854EEMHRD008074 and DST/ICPS/QusT/Theme-3/2019/Q69.
\end{acknowledgments}

\section*{Data Availability Statement}
The data that support the findings of this study are available from the corresponding author upon reasonable request.

\appendix

\section{Derivation of Eq. (\ref{inter})}
\label{Appa}
The initial operators are given by 
\begin{eqnarray}\label{ini}
A=I_x\otimes\mathbb{I}\quad\text{and}\quad B=\mathbb{I}\otimes J_x.  
\end{eqnarray}
First, we consider $A$ and write it in the computational basis as follows:
\begin{eqnarray}\label{eleA}
A&=&\left\{\dfrac{1}{2}\sum_{m, m'=-J}^{J}\left[P(J, m) \delta_{m', m+1}+ Q(J, m) \delta_{m', m-1} \right]|m\rangle\langle m'|\right\}\nonumber\\
&&\hspace{4cm}\otimes \sum_{m''=-J}^{J}|m''\rangle\langle m''|\nonumber\\
&=&\dfrac{1}{2}\sum_{m, m', m''=-J}^{J}\left[P(J, m) \delta_{m', m+1}+ Q(J, m) \delta_{m', m-1} \right] \nonumber\\
&&\hspace{4cm}|mm''\rangle\langle m'm''|, 
\end{eqnarray}
where $P(J, m)=\sqrt{(J-m)(J+m+1)}$, and $Q(J, m)=\sqrt{(J+m)(J-m+1)}$. Now, consider an arbitrary bipartite unitary operator $U$ with the symmetry operator $\hat{F}_z$, i.e., $[U, \hat{F}_z]=0$. Then, $U$ can be decomposed as $\oplus_{F_z}U_{F_z}$, where $U_{F_z}$ represents the unitary contribution having supported over the invariant subspace labeled by the charge $F_z$. Moreover, $F_z$ varies from $-2J$ to $2J$. For an arbitrary $F_z$, the corresponding subspace unitary operator can be written in the computational basis as follows:
\begin{eqnarray}
U_{F_z}=\sum_{\substack{m_1, m_2=-J\\m_1+m_2=F_z}}^{J}\sum_{\substack{m_3, m_4=-J\\m_3+m_4=F_z}}^{J}u_{m_1m_2, m_3m_4} |m_1m_2\rangle\langle m_3m_4|. 
\end{eqnarray}
Then, for some other $F'_{z}$, where $-2J\leq F'_{z}\leq 2J$, it follows that 
\begin{align}
U^{\dagger}_{F_z}A U_{F'_z}=&\sum_{\substack{m_1, m_2=-J\\m_1+m_2=F_z}}^{J}\sum_{\substack{m_3, m_4=-J\\m_3+m_4=F_z}}^{J}\sum_{\substack{m'_1, m'_2=-J\\m'_1+m'_2=F'_z}}^{J}\sum_{\substack{m'_3, m'_4=-J\\m'_3+m'_4=F'_z}}^{J}\nonumber\\
&u^*_{m_1m_2, m_3m_4}u_{m'_1m'_2, m'_3m'_4}\nonumber\\
&|m_3m_4\rangle\langle m_1m_2|A|m'_1m'_2\rangle\langle m'_3m'_4|, 
\end{align}
where the elements $\langle m_1m_2|A|m'_1m'_2\rangle$ follow Eq. (\ref{eleA}) and can be written explicitly as follows:
\begin{eqnarray}
\langle m_1m_2|A|m'_1m'_2\rangle&=&\dfrac{1}{2} \left[P(J, m_1) \delta_{m'_1, m_1+1}+ Q(J, m_1) \delta_{m'_1, m_1-1} \right]\nonumber\\
&&\hspace{4cm}\delta_{m_2m'_2}.
\end{eqnarray}
By imposing the condition that $m_1+m_2=F_z$ and $m'_1+m'_2=F'_z$, one can see that $\langle m_1m_2|A|m'_1m'_2\rangle$ returns a non-zero value only when $F_z=F'_z\pm 1$. As a result, the operator $U^{\dagger}_{F_z}A U_{F'_z}$ returns a non-zero matrix only when $F_z= F'_z\pm 1$. A similar reasoning can be extended to the operator $B$, leading to the following:
\begin{eqnarray}
\left. 
  \begin{array}{ c l }
U_{F_z}^{\dagger}AU_{F'_z}\neq\mathbf{0}\\
U_{F_z}^{\dagger}BU_{F'_z}\neq\mathbf{0}
\end{array}
\right\}\quad\text{iff}\quad F'_z=F_z\pm 1, 
\end{eqnarray}
where $\mathbf{0}$ indicates the null matrix or zero matrix, which contains only zeros. The above equation holds for any arbitrary bipartite unitary operator conserving the total magnetization $F_z$ as long as the initial operator $A$ and $B$ are chosen according to Eq. (\ref{ini}).

\section{Derivation of Eqs. (\ref{23}) and (\ref{24})}
\label{AppB}
We first compute the two point correlator $\Tr\left( U^{\dagger}A^2UB^2 \right)$. Recall that $U=\oplus_{\substack{F_{z}}}U_{F_z}$, where $F_z$ varies from $-2J$ to $+2J$. The unitaries $U_{F_z}$s have non-trivial support over the invariant subspaces labeled by the quantum number $F_z$. To evaluate the RMT value, we choose these unitaries randomly from the circular orthogonal ensemble. We now write 
\begin{eqnarray}
\Tr\left( U^{\dagger}A^2UB^2 \right)&=&\Tr\left[\left(\underset{F_z}{\oplus} U_{F_z}\right)^{\dagger}A^2\left(\underset{F'_z}{\oplus}U_{F'_z}\right)B^2 \right] \nonumber\\
&=&\sum_{F_z, F'_z=-2J}^{2J}\Tr\left( U^{\dagger}_{F_z}A^2U_{F'_z}B^2 \right)\nonumber\\
&=&\sum_{F_z=-2J}^{2J}\Tr\left( U^{\dagger}_{F_z}A^2U_{F_z} B^2\right). 
\end{eqnarray}
The third equality follows directly from Eq. (\ref{inter}). We now compute the RMT value of the two-point correlator by averaging the two-point correlator over all the subspace unitaries $U_{F_z}\in \text{COE}(2J+1-|F_z|)$. 
\begin{eqnarray}
\overline{C_{2}}=\dfrac{1}{(2J+1)^2}\sum_{F_z=-2J}^{2J}\Tr\left( \overline{U^{\dagger}_{F_z}A^2U_{F_z}} B^2 \right). 
\end{eqnarray}
For an arbitrary matrix $P$ and a random $W\in \text{COE} (d)$, the operator $\overline{W^{\dagger}PW}$ can be evaluated as \cite{brouwer1996diagrammatic} 
\begin{eqnarray}
\overline{W^{\dagger}PW}=\dfrac{1}{d+1}\left( P^{T}+\Tr(P)\mathbb{I} \right),  
\end{eqnarray}
where $d$ is the dimension of the Hilbert space, and $P^T$ denotes the transpose matrix of $P$. In the above expression, all the operators have been assumed to have equal dimensions. We now replace $P$ with $A^2$ and consider the average over the subspace unitaries belonging to the COE. In this case, the above equation gets modified as follows:
\begin{eqnarray}
\overline{U^{\dagger}_{F_z}A^2U_{F_z}} = \dfrac{1}{2J+2-|F_z|}\left[[A^2]^{T}_{F_z}+\Tr\left( [A^2]_{F_z}\right)\mathbb{I}_{F_z} \right], 
\end{eqnarray}
where $[A^2]_{F_z}$ denotes the projection of $A^2$ onto the invariant subspace with the quantum number $F_z$ and $\mathbb{I}_{F_z}$ is the identity operator supported non-trivially over the same subspace. It then follows that 
\begin{eqnarray}
\overline{C_{2}}=\dfrac{1}{(2J+1)^2}\sum_{F_z}\dfrac{\Tr([A^2]_{F_z}^TB^2)+\Tr([A^2]_{F_z})\Tr([B^2]_{F_z})}{2J+2-|F_z|}.\nonumber\\ 
\end{eqnarray}
We now show that the four-point correlator, on the other hand, vanishes for the $\hat{I}_x\hat{J}_x$-OTOC. To do so, we consider 
\begin{align}
\overline{C_4}=&\dfrac{1}{(2J+1)^2}\nonumber\\
&\Tr\left[\overline{\left(\underset{F^{a}_z}{\oplus}U_{F^{a}_z}\right)^{\dagger}A\left(\underset{F^{b}_z}{\oplus}U_{F^{b}_z}\right)B\left(\underset{F^{c}_z}{\oplus}U_{F^{c}_z}\right)^{\dagger}A\left(\underset{F^{d}_z}{\oplus}U_{F^{d}_z}\right)B}\right] \nonumber\\
=&\dfrac{1}{(2J+1)^2}\sum_{F^{a}_zF^{b}_zF^{c}_zF^{d}_z}\Tr\left(\overline{U^{\dagger}_{F^{a}_z}AU_{F^{b}_z}BU^{\dagger}_{F^{c}_z}AU_{F^{d}_z}}B\right).
\end{align}
Equation (\ref{inter}) implies that the four-point correlator vanishes unless $F^{a}_{z}=F^{b}_{z}\pm 1$, and $F^{c}_{z}=F^{d}_{z}\pm 1$. Hence, 
\begin{eqnarray}
\overline{C_{4}}=\dfrac{1}{(2J+1)^2}\sum_{F^{a}_z, F^{c}_{z}}\Tr\left(\overline{U^{\dagger}_{F^{a}_z}AU_{F^{a}_z\pm 1}BU^{\dagger}_{F^{c}_z}AU_{F^{c}_z\pm 1}}B  \right).
\end{eqnarray}
Interestingly, the above expression can be further simplified by noting that the terms inside the summation vanish unless $F^{a}_{z}=F^{c}_{z}$. It then follows that 
\begin{eqnarray}
\overline{C_{4}}=\dfrac{1}{(2J+1)^2}\sum_{F_{z}=-2J}^{2J}\Tr\left(\overline{U^{\dagger}_{F_z}AU_{F_z\pm 1}BU^{\dagger}_{F_z}AU_{F\pm 1}}B\right).    
\end{eqnarray}
Here, we replaced $F^{a}_z$ with $F_z$. A few remarks are in order. Firstly, the subspace operators $U_{F_z}$ and $U_{F_z\pm 1}$ act non-trivially on the invariant subspaces with the total magnetization $F_z$ and $F_z\pm 1$, respectively. Hence, the COE average over the unitaries $U_{F_z}$ and $U_{F_z\pm 1}$ should be performed independently. See Ref. \cite{brouwer1996diagrammatic} for more details concerning the averages over COE matrices. Upon performing the average, we get
\begin{eqnarray}
\overline{C_{4}}=0.    
\end{eqnarray}
Therefore, $C_{\text{RMT}}=\overline{C_{2}}$.

\bibliographystyle{unsrt}
\bibliography{reference_otoc}

\end{document}